\documentclass[12pt,epsf]{article}
\usepackage[pdftex]{graphicx}
\pdfoutput=1
\usepackage[hang,small,bf]{caption}
\usepackage[subrefformat=parens]{subcaption}
\usepackage{amsmath,amssymb}
\usepackage{bm}
\usepackage{ascmac}
\usepackage{braket}
\graphicspath{{./fig/}}
\usepackage{comment}
\usepackage{cite}

\newcommand{\beq}{\begin{equation}}
\newcommand{\eeq}{\end{equation}}
\newcommand{\bea}{\begin{eqnarray}}
\newcommand{\eea}{\end{eqnarray}}

\newcommand{\ba}{\begin{aligned}}
\newcommand{\ea}{\end{aligned}}

\def\pe2{p_E^2}

\begin{document}
\setlength{\baselineskip}{0.7cm}
\begin{titlepage}
%%%%% PREPRINT NUMBERS %%%%%%
\begin{flushright}
%OMU-PHYS 564  \\
NITEP 137
\end{flushright}
\vspace*{10mm}%
%%%%%%%%%%%%%%%%%%% TITLE %%%%%%%%%%%%%%%%%%
\begin{center}{\Large\bf
Gauge Symmetry Breaking %\\
in Flux Compactification \\
\vspace*{2mm}
with Wilson-line Scalar Condensate
}
\end{center}
%%%%%%%%%%%%%%%% AUTHORS %%%%%%%%%%%%%%%%%%%%%%%
\vspace*{10mm}
\begin{center}
{\large Kento Akamatsu}$^{a}$,  %\footnote{E-mail: m21sa001@st.osaka-cu.ac.jp}
{\large Takuya Hirose}$^{a}$ and %\footnote{E-mail: t-hirose555@ty.osaka-cu.ac.jp}
{\large Nobuhito Maru}$^{a,b}$, %\footnote{E-mail: nmaru@sci.osaka-cu.ac.jp}
\end{center}
%%%%%%%%%%%%%%%%%%%%%%% AFFILIATION %%%%%%%%%%%%
\vspace*{0.2cm}
\begin{center}
%\small
${}^{a}${\it
Department of Physics, Osaka Metropolitan University, \\
Osaka 558-8585, Japan}
\\
%\\[0.2cm]
${}^{b}${\it Nambu Yoichiro Institute of Theoretical and Experimental Physics (NITEP), \\
Osaka Metropolitan University,
Osaka 558-8585, Japan}
%%%%%
\end{center}
%%%%%%%%%%%%%%%%%% ABSTRACT %%%%%%%%%%%%%%%
\vspace*{1cm}

\begin{abstract}
We discuss the gauge symmetry breaking of six dimensional theories in flux compactification 
 with a magnetic flux background and a constant vacuum expectation value (VEV) 
 for the scalar fields, which are zero modes of extra spatial components of the gauge field.
Although the effective potential for the scalar fields are known not to be generated classically and radiatively 
 in a magnetic flux background only, the one-loop effective potential is shown to be generated by the effects of the non-zero constant VEV.   
As illustrations, we calculate the one-loop effective potential in SU(2) and SU(3) Yang-Mills theories. 
In both cases, we expect that the potential minimum is located at non-zero VEV 
 and the gauge symmetry breaking takes place. 
\end{abstract}
\end{titlepage}

%%%%%%%%%%%%%%%%%%%%%%
\section{Introduction}
%%%%%%%%%%%%%%%%%%%%%%
Although the Standard Model (SM) has a successful theory, it still has some problems. 
Many attractive scenarios based on the higher dimensional theory have been proposed 
 as the physics beyond the SM.  
%It has been proposed that some sort of higher dimensional theory %\cite{Manton, Hosotani1, Hosotani2, HIL} 
% is an attractive theory as physics beyond the SM. %solves some problems in the SM. 
In particular, flux compactification, which has been studied in string theory \cite{BKLS,IU}, 
 has many attractive aspects: explanation of the generation number of the SM fermions \cite{Witten,ACKO}
and computation of Yukawa coupling \cite{CIM,0903,highermode,MS}.

Recently, it has been considered that the quantum corrections to the masses of zero-mode of the scalar field 
 induced from extra components of higher dimensional gauge field (called as Wilson-line (WL) scalar field) are cancelled 
 \cite{B1,Lee,B2,HM,2-loop} and are finite \cite{finite}.
The reason why the quantum corrections are cancelled is 
 that the shift symmetry from translation in extra spaces forbids the mass term of scalar field %In other words, 
since the zero-mode of the scalar field can be identified with Nambu-Goldstone (NG) boson 
 of spontaneously broken translational symmetry (or with pseudo-NG boson in \cite{finite}).
This cancellation mechanism may be applied to the hierarchy problem in the SM
 which is the problem that the quantum corrections to the mass of Higgs field 
 are sensitive to the square of the ultraviolet cutoff scale of the theory.
If we regard the Higgs field as the WL scalar field, 
 which is an idea of gauge-Higgs unification \cite{Manton, Hosotani1, Hosotani2, HIL},
 the quantum corrections to the mass of Higgs field is cancelled as mentioned above.   
In the gauge-Higgs unification, 
 the finite Higgs mass is generated by the quantum corrections \cite{HIL, ABQ, LMH, MaruYamashita, HMTY} 
 controlled by the compactification scale. 
If the compactification scale is increased by the absence of the new physics discovery, 
 the fine-tuning problem in the Higgs mass parameter is reintroduced. 
In flux compactification, however, if the translational symmetry in extra spaces is explicitly broken 
 around TeV scale independent of the compactification scale, 
 the light Higgs boson mass is radiatively generated. 
This logic is also applied to the potential of the WL scalar field. 
 %ゲージヒッグスの引用は色々引きましたが、丸さんに任せます

In this paper, we investigate the gauge symmetry breaking 
 in a higher dimensional theory in flux compactification with a magnetic flux background 
 and a constant WL scalar vacuum expectation value (VEV).
First, we consider a six dimensional SU(2) Yang-Mills theory compactified on a torus with a magnetic flux and a constant VEV.
%Because of the existence of the constant VEV, we cannot easily derive Kaluza-Klein (KK) masses.
%However, we can calculate the KK masses in terms of perturbation theory in quantum mechanics,
%and the KK mass can be obtained as a sum of the KK mass induced from flux background and the constant VEV part. %sumとすると少し誤解を招くかも？
Calculating the KK mass spectrum in the presence of both flux background and the constant VEV, 
 we obtain the one-loop effective potential for the WL scalar field.   
%After finding the KK mass, we analyze the effective potential.
Although the effective potential in the flux background only is not radiatively generated, 
 the effective potential in both the flux background and the constant VEV is generated at one-loop 
 and the potential minimum at nonvanishing constant VEV is expected.  
This concludes that gauge symmetry SU(2) is completely broken. 

Next, we consider a six dimensional SU(3) Yang-Mills theory compactified on a torus with a magnetic flux and a constant VEV.
%In general, flux background and the constant VEV can be introduced in the adjoint direction: $N^2-1$ ways.
In this case, we consider two types of configuration where the flux background and the constant VEV are developed.  
One is that the flux background and the constant VEV are in the eighth and the first components of SU(3).  
The other is that the flux background and the constant VEV are in the eighth and the sixth components of SU(3). 
Calculating the one-loop effective potential similarly as is done in SU(2) Yang-Mills theory, 
  we find (expect) that the potential is minimized at nonvanishing constant VEV in the former (latter) case. 
In the former (latter) case, the gauge symmetry breaking SU(3) $\to$ U(1) $\times$ U(1)(SU(3) $\to$ U(1)) is found, respectively. 
%One can easily calculate the KK masses d the other cannot easily compute the KK masses as SU(2) case. 
%We show the effective potential in each case.
%From KK mass of gauge field, we read the symmetry breaking pattern.

This paper is organized as follows.
We give a setup of a six-dimensional SU(2) Yang-Mills theory with magnetic flux compactification and introduce the constant VEV 
%After deriving the mass of fields, we analyze the effective potential in our model 
 in section \ref{SU(2)}.
We furthermore consider a six-dimensional SU(3) Yang-Mills theory with magnetic flux compactification in section 3, 
 where two types of configuration for the flux background and the constant VEV to be taken. %\ref{SU(3)dir1,8}.
In both sections, the one-loop effective potential for the WL scalar field is calculated 
 and the gauge symmetry breaking is discussed.  
%In this section, we introduce the constant VEV in the direction of gauge index $a=1$ and the flux background in the direction of gauge index $a=8$.
%We also see that the effective potential has a periodicity and the gauge symmetry is broken.
%In section \ref{SU(3)dir6,8}, we consider a six-dimensional SU(3) Yang-Mills theory with magnetic flux compactification 
% introducing the constant VEV in the direction of gauge index $a=6$ and the flux background in the direction of gauge index $a=8$.
%This section is similar to section \ref{SU(2)}, and we apply the result in section \ref{SU(2)}.
In the last section, we devote our summary. 
In Appendix A, the calculation of the KK mass spectrum at the second order perturbation is summarized.  

%%%%%%%%%%%%%%%%%%%%%%
\section{SU(2) Yang-Mills theory}
\label{SU(2)}
%%%%%%%%%%%%%%%%%%%%%%
We consider a six-dimensional SU(2) Yang-Mills theory with two nontrivial backgrounds: 
 a constant magnetic flux background and an ordinary constant vacuum expectation value.
%%%%%%%%%%%%%%%%%%%%%%
\subsection{Set up}
%%%%%%%%%%%%%%%%%%%%%%
Six-dimensional spacetime is $M^4\times T^2$, where $M^4$ is a Minkowski spacetime and $T^2$ is a two-dimensional square torus.
The Lagrangian of SU(2) Yang-Mills theory in six dimensions is
	\begin{align}
	\label{Lag6}
	\mathcal{L}_6&=-\frac{1}{4}F^a_{MN}F^{aMN}\nonumber \\
	&=-\frac{1}{4}F^a_{\mu\nu}F^{a\mu\nu}-\frac{1}{2}F^a_{\mu5}F^{a\mu5}
		-\frac{1}{2}F^a_{\mu6}F^{a\mu6}-\frac{1}{2}F^a_{56}F^{a56},
	\end{align}
where the field strength tensor and the covariant derivative are defined by
	\begin{align}
	F^a_{MN}&=\partial_MA^a_N-\partial_NA^a_M-ig[A_M,A_N]^a,\\
	D_MA^a_N&=D_M^{ac}A_N^c
		=(\delta^{ac}\partial_M+g\epsilon^{abc}A^b_M)A^c_N\nonumber \\
	&=\partial_MA^a_N-ig[A_M,A_N]^a.
	\end{align}
The spacetime indices are $M,N=0,1,2,3,5,6, \mu,\nu=0,1,2,3, m,n=5,6$, and the gauge indices are $a,b,c=1,2,3$.
The metric convention $\eta_{MN}=\text{diag}(-1,+1,\cdots,+1)$ is employed.
$\epsilon^{abc}$ is a totally anti-symmetric tensor of SU(2).

%First, 
We discuss how the two backgrounds are introduced in our model.
First, the constant magnetic flux, is given by the VEV of the fifth and the sixth component of the gauge fields, 
 $A^3_{5,6}$, which must satisfy their classical equation of motion:
	\begin{align}
	D^m\braket{F^a_{mn}}=0.
	\end{align}
Second, the ordinary constant background is generated 
 by the quantum correction in the sixth component of the gauge field $A^1_6$, for simplicity
 \footnote{In general, the constant background can be introduced by the fifth component of the gauge field $A^1_5$. }.
%Note that this constant background does not satisfy the ``classical'' equation of motion above, 
% but satisfies if the one-loop potential is taken into account. 
%: it is treated as zero in the equation.
In this section, we choose a solution
	\begin{align}
	\braket{A^1_6}=v,~~~\braket{A^3_5}=-\frac{1}{2}fx_6,~~~
		\braket{A^3_6}=\frac{1}{2}fx_5~~
		\text{and}~~\braket{A^{1,2}_5}=\braket{A^2_6}=0.
	\end{align}
$\braket{A^3_{5,6}}$ introduce a magnetic field parametrized by a constant $f$, %which satisfies 
 namely $\braket{F^3_{56}}=f$.
Note that the flux background spontaneously breaks a translational invariance on the torus.
The flux background breaks the gauge symmetry, which is broken to U(1) in this case.
The flux is also associated with the degeneracy:
	\begin{align}
	\frac{g}{2\pi}\int_{T^2}dx_5dx_6\braket{F^3_{56}}=\frac{g}{2\pi}L^2f
		=N\in\mathbb{Z},
	\end{align}
where $L^2$ is an area of the torus. 
For simplicity, we set $L=1$ from now on.
It is useful to define $\partial$ and the scalar fields $\phi^a$ as
	\begin{align}
	\label{cmpl}
	\partial\equiv\partial_z=\partial_5-i\partial_6,~~~
		z\equiv\frac{1}{2}(x_5+ix_6),~~~\phi^a=\frac{1}{\sqrt{2}}(A^a_6+iA^a_5).
	\end{align}
In these complex coordinates, the VEVs of $\phi^{1,3}$ are given by
	\begin{align}
	\braket{\phi^1}=\frac{v}{\sqrt{2}}, \qquad \braket{\phi^3}=\frac{f\bar z}{\sqrt{2}},
	\end{align}
and we expand $\phi^a$ around the backgrounds:
	\begin{align}
	\phi^a=\braket{\phi^a}+\varphi^a,
	\end{align}
where $\varphi^a$ are quantum fluctuations.

The Lagrangian \eqref{Lag6} can be rewritten by using the new coordinates \eqref{cmpl} as follows:
	\begin{align}
	\label{Lag6p}
	\mathcal{L}_6=&-\frac{1}{4}F^a_{\mu\nu}F^{a\mu\nu}
		-\partial_\mu\bar\phi^a\partial^\mu\phi^a
		-\frac{1}{2}DA^a_\mu\overline{D}A^{a\mu}\nonumber \\[2mm]
	&-\frac{i}{\sqrt{2}}(\partial_\mu\phi^a\bar\partial A^{a\mu}
			-\partial_\mu\bar\phi^a\partial A^{a\mu})\nonumber \\[2mm]
	&+ig\left(\partial_\mu\phi^a[A^\mu,\bar\phi]^a
		+\partial^\mu\bar\phi^a[A_\mu,\phi]^a\right)\nonumber \\[2mm]
	&-\frac{1}{4}\left(D\bar\phi^a+\overline{D}\phi^a+
		\sqrt{2}g[\phi,\bar\phi]^a\right)^2,
	\end{align}
where
	\begin{align}
	\begin{cases}
	D\Phi^a\equiv(D_5-iD_6)\Phi^a=\partial\Phi^a-\sqrt{2}g[\phi,\Phi]^a,
		\\[2mm]
	\overline{D}\Phi^a\equiv(D_5+iD_6)\Phi^a=\bar{\partial}\Phi^a
		+\sqrt{2}g[\bar{\phi},\Phi]^a,
	\end{cases}
	\end{align}
which express the covariant derivatives with respect to the complex coordinates in the torus.
$\Phi^a$ denote arbitrary fields in the adjoint representation.
We can remove the mixing terms between the gauge and the scalar fields 
 %which is 
 in the second line of eq. \eqref{Lag6p} by introducing the gauge-fixing terms with a gauge parameter $\xi$:
	\begin{align}
	\mathcal{L}_{g-f}\equiv&-\frac{1}{2\xi}
		(D_\mu A^{a\mu}+\xi\mathcal{D}_mA^{am})^2\nonumber \\[2mm]
	=&-\frac{1}{2\xi}D_\mu A^{a\mu}D_\nu A^{a\nu}
		-\frac{g}{\sqrt{2}}\left(\partial\bar{\phi}^a[A_\mu,A^\mu]^a
			-\bar\partial\phi^a[A_\mu,A^\mu]^a\right)\nonumber \\[2mm]
	&+\frac{\xi}{4}(\mathcal{D}\bar\phi^a
		-\overline{\mathcal{D}}\phi^a)^2
		+\frac{i}{\sqrt{2}}(\partial_\mu\phi^a\bar\partial A^{a\mu}
		-\partial_\mu\bar\phi^a\partial A^{a\mu}).
	\end{align}
The new covariant derivatives $\mathcal{D},\overline{\mathcal{D}}$ are defined 
 by replacing $\phi^a,\bar\phi^a$ in $D,\overline{D}$ with the VEVs $\braket{\phi^a},\braket{\bar\phi^a}$, respectively. 
 Due to the gauge-fixing, we need to introduce the Faddeev-Popov ghost fields and their Lagrangian is
	\begin{align}
	\mathcal{L}_{ghost}=-\bar c^a(D_\mu D^\mu+\xi D_m\mathcal{D}^m)
		c^a.
	\end{align}
Then, the total Lagrangian is given by 
	\begin{align}
	\mathcal{L}_{total}=
	&-\frac{1}{4}F^a_{\mu\nu}F^{a\mu\nu}
		-\frac{1}{2\xi}D_\mu A^{a\mu}D_\nu A^{a\nu}
		-\partial_\mu\bar\phi^a\partial^\mu\phi^a
		-\frac{1}{2}DA^a_\mu\overline{D}A^{a\mu}\nonumber \\[2mm]
	&-\frac{g}{\sqrt{2}}\left(\partial\bar{\phi}^a[A_\mu,A^\mu]^a
			-\bar\partial\phi^a[A_\mu,A^\mu]^a\right)
		+\frac{\xi}{4}(\mathcal{D}\bar\phi^a
		-\overline{\mathcal{D}}\phi^a)^2\nonumber \\[2mm]
	&+ig\left(\partial_\mu\phi^a[A^\mu,\bar\phi]^a
		+\partial^\mu\bar\phi^a[A_\mu,\phi]^a\right)\nonumber \\[2mm]
	&-\frac{1}{4}\left(D\bar\phi^a+\overline{D}\phi^a+
		\sqrt{2}g[\phi,\bar\phi]^a\right)^2-\bar c^a(D_\mu D^\mu+\xi D_m\mathcal{D}^m)c^a.
	\end{align}
%Second, we discuss mass eigenstates and eigenvalues of the fields $A^a_\mu,\varphi^a,c^a$, 
%which will reminds us of a discussion of the Landau level in quantum mechanics.
For simplicity, we choose the Feynman gauge $\xi=1$ throughout this paper.

%%%%%%%%%%%%%%%%%%%%%%
\subsection{The mass of the gauge fields $A^a_\mu$}
\label{SU(2)A}
%%%%%%%%%%%%%%%%%%%%%%
We will discuss mass eigenstates and eigenvalues of the fields $A^a_\mu,\varphi^a,c^a$.
%Here, 
In this subsection, we find mass eigenvalues and eigenstates of the gauge fields $A^a_\mu$.
The mass term of gauge field corresponds to the background part of $-DA^a_\mu\overline{D}A^{a\mu}/2$:
	\begin{align}
	\mathcal{L}_{AA}=-\frac{1}{2}\mathcal{D}A^a_\mu\overline{\mathcal{D}}A^{a\mu}
	=-\frac{1}{2}A^a_\mu\left[-\mathcal{D}\overline{\mathcal{D}}\hspace{0.5mm}\right]A^{a\mu}.
	\end{align}
We would like to regard the background covariant derivatives $\mathcal D,\overline{\mathcal D}$ 
 as creation and annihilation operators, respectively. %: their commutation relation needs to be just a constant matrix.
In a matrix form, they are expressed as
	\begin{align}
	\mathcal{D}=\left[\begin{array}{ccc}
	\partial & igf\bar{z} & 0  \\
	-igf\bar{z} & \partial & igv  \\
	0 & -igv & \partial 
	\end{array}\right], \qquad
	\overline{\mathcal{D}}=\left[\begin{array}{ccc}
	\bar{\partial} & -igfz & 0 \\
	igfz & \bar{\partial} & -igv  \\
	0 & igv & \bar{\partial}
	\end{array}\right].
	\label{covdiag}
	\end{align}
Diagonalizing them, we obtain
	\begin{align}
	\begin{cases}
	\mathcal{D}_{diag}=\text{diag}\left(\partial,\partial+g\sqrt{f^2\bar z^2+v^2},\partial-g\sqrt{f^2\bar z^2+v^2}\right), 
	\\[2mm]
	\overline{\mathcal D}_{diag}=\text{diag}\left(\bar\partial,\bar\partial-g\sqrt{f^2z^2+v^2},
		\bar\partial+g\sqrt{f^2z^2+v^2}\right).
	\end{cases}
	\label{diag:sec2}
	\end{align}
Their commutation relations is
	\begin{align}
	\left[\overline{\mathcal D}_{diag},\mathcal{D}_{diag}\right]^{ac}=
	gf^2\left(\frac{z}{\sqrt{f^2z^2+v^2}}+\frac{\bar z}{\sqrt{f^2\bar z^2+v^2}}\right)\left[\begin{array}{ccc}
	0 & 0 & 0 \\
	0 & 1 & 0 \\
	0 & 0 & -1
	\end{array}\right],
	\end{align}
which depends on extra space coordinates. 
Therefore, $\mathcal D,\overline{\mathcal D}$ cannot be identified with creation and annihilation operators. 
%and we cannot use the method of the Landau level. 

%By the way,
%In general, it is said that Kaluza-Klein (KK) mass spectrum in higher dimensional theory is based on idea of quantum mechanics.
%From this perspective, we can easily calculate the mass of gauge field when the constant VEV $v$ is zero,
%and then we try to apply the perturbation theory to our model whose $v$ is much smaller than $1$.
Since Kaluza-Klein (KK) mass spectrum cannot be exactly solved 
 by using the creation and annihilation operators, 
 we would like to find them perturbatively by the expansion in $v$. 
In this expansion, $vL \ll 1$ or $v \ll 1$ in the present case is assumed. 
In other words, we consider the case where the compactification scale is much larger than the constant VEV $v$. 
From \eqref{covdiag}, we define the unperturbed parts $\mathcal D_3,\overline{\mathcal D}_3$ 
 and the perturbed part $V$ as
	\begin{align}
	\mathcal{D}_3\equiv\left[\begin{array}{ccc}
	\partial & igf\bar{z} & 0  \\
	-igf\bar{z} & \partial & 0  \\
	0 & 0 & \partial 
	\end{array}\right],~~~
	\overline{\mathcal{D}}_3\equiv\left[\begin{array}{ccc}
	\bar{\partial} & -igfz & 0 \\
	igfz & \bar{\partial} & 0  \\
	0 & 0 & \bar{\partial}
	\end{array}\right]
	\label{covSU(2)}
	\end{align}
and
	\begin{align}
	V\equiv\left[\begin{array}{ccc}
	0 & 0 & 0 \\
	0 & 0 & igv  \\
	0 & -igv & 0
	\end{array}\right],
	\label{perturSU(2)}
	\end{align}
respectively.
In these notations, the covariant derivatives can be expressed as $\mathcal{D}=\mathcal{D}_3+V$ 
 and $\overline{\mathcal{D}}=\overline{\mathcal{D}}_3+\overline{V}$.
$\mathcal D_3$ and $\overline{\mathcal D}_3$ can be identified with creation and annihilation operators, 
 which are diagonalized as follows. 
%We diagonalize them and obtain the diagonalized covariant derivatives $\mathcal D_{3,diag},\overline{\mathcal D}_{3,diag}$:
	\begin{align}
	\begin{cases}
	\mathcal D_{3,diag}=\text{diag}(\partial-gf\bar{z},
		\partial+gf\bar{z},\partial), 
		\\[2mm]
	\overline{\mathcal D}_{3,diag}=\text{diag}(\bar\partial+gfz,
		\bar\partial-gfz,\bar\partial).
	\end{cases}
	\end{align}
Their diagonalizing unitary matrix $U_3$, which satisfies 
$U_3^{-1}\mathcal D_3U_3=\mathcal D_{3,diag}$ and $U_3^{-1}
\overline{\mathcal D}_3U_3=\overline{\mathcal D}_{3,diag}$,
is
	\begin{align}
	U_3=\frac{1}{\sqrt{2}}\left[\begin{array}{ccc}
	1 & i & 0 \\
	i & 1 & 0 \\
	0 & 0 & \sqrt{2}
	\end{array}\right].
	\end{align}
The commutation relation between $\mathcal D_{3,diag}$ and $\overline{\mathcal D}_{3,diag}$ is
	\begin{align}
	\left[i\overline{\mathcal D}_{3,diag},i\mathcal D_{3,diag}\right]^{ac}=
	2gf\left[\begin{array}{ccc}
		1 & 0 & 0 \\
		0 & -1 & 0 \\
		0 & 0 & 0 
	\end{array}\right].
	\label{commrel}
	%\equiv-2gf[a,a^\dagger]^{ac},
	\end{align}
The creation and annihilation operators are defined as
%where the creation and annihilation operators are
	\begin{align}
	a=\frac{i}{\sqrt{\alpha_2}}\overline{\mathcal D}_{3,diag}, \qquad 
	a^\dagger=\frac{i}{\sqrt{\alpha_2}}\mathcal D_{3,diag},
	\end{align}
%from \eqref{commrel}.Here, we denote 
where $\alpha_2\equiv2gf$.
The components of the creation and annihilation operators are summarized as follows:
	\begin{align}
	\begin{cases}
	\displaystyle a_1\equiv\frac{i}{\sqrt{ \alpha_2}}(\bar\partial+gfz), \\[4mm]
	\displaystyle a_2\equiv\frac{i}{\sqrt{ \alpha_2}}(\bar\partial-gfz), \\[4mm]
	\displaystyle a_3\equiv\frac{i}{\sqrt{ \alpha_2}}\bar\partial,
	\end{cases}
	~~\text{and}~~~~
	\begin{cases}
	\displaystyle a^\dagger_1\equiv\frac{i}{\sqrt{ \alpha_2}}(\partial-gf\bar z), 
		\\[4mm]
	\displaystyle a^\dagger_2\equiv\frac{i}{\sqrt{ \alpha_2}}(\partial+gf\bar z), 
		\\[4mm]
	\displaystyle a^\dagger_3\equiv\frac{i}{\sqrt{ \alpha_2}}\partial.
	\end{cases}
	\end{align}
We note that $a_3$ and $a^\dagger_3$ have no flux effects and play no role of creation and annihilation operators. 
$a^\dagger_1$ and $a_2$ are creation operators and $a_1$ and $a^\dagger_2$ are annihilation operators:
the roles of creation and annihilation operators for $a_2$ and $a^\dagger_2$ are inverted due to the commutation relation for the $2$-direction $[a_2,a^\dagger_2]=-1$.
The ground state mode functions are determined by
	\begin{align}
	a_1\xi_{0,j}=0, \qquad a^\dagger_2\bar\xi_{0,j}=0,
	\end{align}
where $\xi_{0,j}$ are the functions of $z$ (in detail, see \cite{CIM}) %:
%	\begin{align}
%	\xi_{0,j}\equiv\xi_{0,j}(z)=\mathcal{N}e^{i\pi|N|z\Im{z}}\cdot
%		\vartheta\left[\begin{array}{c}j/|N| \\ 0 \end{array}\right](|N|z,|N|i),
%	\end{align}
%where $\mathcal N$, $|N|$ and $\vartheta$ are a normalization constant, 
%a degeneracy number and the theta function with characteristics, respectively.
and $j$ labels the degeneracy of the ground state: $j=0,\cdots,|N|-1$.
Higher mode functions $\xi_{n,j}$ are constructed similar to the harmonic oscillator case \cite{highermode}:
	\begin{align}
	\xi_{n_1,j}=\frac{1}{\sqrt{n_1!}}(a^\dagger_1)^{n_1}\xi_{0,j},~~
	\bar\xi_{n_2,j}=\frac{1}{\sqrt{n_2!}}(a_2)^{n_2}\bar\xi_{0,j}
	~~(n_{1,2}\in\mathbb{Z},~n_{1,2}\geq0).
	\end{align}
The functions $\xi_{n,j}$ satisfy the orthogonal conditions
	\begin{align}
	\int_{T^2}d^2x~\bar\xi_{n,j}~\xi_{n^\prime,j^\prime}
		=\delta_{n,n^\prime}\delta_{j,j^\prime}.
	\end{align}
To be operated by $a$ or $a^\dagger$, we should define the states with the gauge indices:
	\begin{align}
	\psi^1_{n_1,j}\equiv\left[\begin{array}{c}
	\xi_{n_1,j} \\ 0 \\ 0 \end{array}\right], \qquad 
%	~~\text{and}~~
	\psi^2_{n_2,j}\equiv\left[\begin{array}{c}
	0 \\ \bar\xi_{n_2,j} \\ 0 \end{array}\right].
	\end{align}
Moreover, using the periodic boundary condition of torus, we define the eigenstates for the 3-direction:
	\begin{align}
	\psi^3_{l,m}\equiv\left[\begin{array}{c}
	0 \\ 0 \\ \lambda_{l,m} \end{array}\right]~~~(l,m\in\mathbb{Z}),
	\end{align}
where $\lambda_{l,m}$ are the functions of $x_5$ and $x_6$:
	\begin{align}
	\lambda_{l,m}\equiv\lambda_{l,m}(x_5,x_6)
		=\exp[\hspace{0.3mm}2\pi i(lx_5+mx_6)].
	\end{align}
These functions $\lambda_{l,m}$ also satisfy the orthogonal conditions
	\begin{align}
	\int_{T^2}d^2x~\bar\lambda_{l,m}\lambda_{l^\prime,m^\prime}
		=\delta_{l,l^\prime}\delta_{m,m^\prime}.
	\end{align}
%We note that the torus integration of $\xi_{n,j}$ and $\lambda_{l,m}$ are nontrivial 
%and we will find that this integration makes crucial effect in a potential analysis.

The eigenstates satisfy the following relations
	\begin{align}
	\begin{cases}
	a\psi^1_{n_1,j}=\sqrt{n_1}\psi^1_{n_1-1,j}, \\[2mm]
	a\psi^2_{n_2,j}=\sqrt{n_2+1}\psi^2_{n_2+1,j}, \\[2mm]
	\displaystyle a\psi^3_{l,m}\hspace{1.1mm}=-\sqrt{\frac{4\pi^2}{\alpha_2}}(l+im)\psi^3_{l,m},
	\end{cases}
	~~\text{and}~~~
	\begin{cases}
	a^\dagger\psi^1_{n_1,j}=\sqrt{n_1+1}\psi^1_{n_1+1,j}, \\[2mm]
	a^\dagger\psi^2_{n_1,j}=\sqrt{n_2}\psi^2_{n_2-1,j}, \\[2mm]
	\displaystyle a^\dagger\psi^3_{l,m}
		\hspace{1.1mm}=-\sqrt{\frac{4\pi^2}{\alpha_2}}(l-im)\psi^3_{l,m}.
	\end{cases}
	\end{align}
For convenience we unify the labels of the eigenstates:
	\begin{align}
	\psi^1_{\{n_1\}}\equiv\psi^1_{n_1,j},~~\psi^2_{\{n_2\}}\equiv\psi^2_{n_2,j}, 
	\quad 
		%~~\text{and}~~
		\psi^3_{\{n_3\}}\equiv\psi^3_{l,m}.
	\end{align}
Their products of %different directional 
eigenstates in the different directions are zero.
	\begin{align}
	\label{psic}
	\left(\psi^a_{\{n_a\}}\right)^\dagger
		\psi^c_{\{n_c\}}=0~~(a\neq c).
	\end{align}

Then we define the unperturbed Hamiltonian $H_0$ and the perturbation $V_1,V_2$:
	\begin{align}
	-\mathcal{D}\overline{\mathcal{D}}
	=&~U_3\left[-\mathcal D_{3,diag}\overline{\mathcal D}_{3,diag}
		+\mathcal D_{3,diag}U_3^{-1}VU_3\right.\nonumber \\
	&\hspace{1cm}\left.-U_3^{-1}VU_3\overline{\mathcal D}_{3,diag}
		+(U_3^{-1}VU_3)^2\right]U_3^{-1}\nonumber \\
	\equiv&~U_3\left[H_0+V_1+V_1^\dagger+V_2\right]U_3^{-1}\nonumber \\
	\equiv&~U_3HU_3^{-1}.
	\label{DDbartransform}
	\end{align}
From the previous discussion, $H_0$ is expressed as
	\begin{align}
	\label{H0}
	H_0=\left[\begin{array}{ccc}
	\alpha_2 n_1 & 0 & 0 \\
	0 & \alpha_2(n_2+1) & 0 \\
	0 & 0 & 4\pi^2(l^2+m^2)
	\end{array}\right],
	\end{align}
and its eigenstates of the gauge fields are defined by
	\begin{align}
	A^{a^\prime}_\mu U_3^{a^\prime a}\equiv\widetilde{A}^a_\mu
		\equiv\sum_{\{n_a\}}\widetilde{A}^a_{\mu,\{n_a\}}\psi^a_{\{n_a\}}, 
		\qquad 
%	~~\text{and}~~
	\left(U^{-1}_3\right)^{aa^\prime}A^{a^\prime\mu}\equiv\widetilde{A}^{a\mu}
		\equiv\sum_{\{n_a\}}\widetilde{A}^{a\mu}_{\{n_a\}}\psi^a_{\{n_a\}}.
	\end{align}
The perturbation $V_1$ and $V_2$ are
	\begin{align}
	V_1=\frac{gv}{\sqrt{2}}\left[\begin{array}{ccc}
	0 & 0 & \partial-gf\bar{z} \\
	0 & 0 & i(\partial+gf\bar{z}) \\
	\partial & -i\partial & 0
	\end{array}\right], 
	\qquad 
%	~~\text{and}~~~
	V_2=\frac{g^2v^2}{2}\left[\begin{array}{ccc}
	1 & -i & 0 \\
	i & 1 & 0 \\
	0 & 0 & 2
	\end{array}\right],
	\end{align}
respectively.
As was seen from the unperturbed Hamiltonian \eqref{H0}, we find that there are degeneracy: 
 $\psi^1_{0,j}$ and $\psi^3_{l,m}$, $\psi^1_{n+1,j}$ and $\psi^2_{n,j}$. 
We thus should be careful to calculate their energies in perturbation. 

For $\psi^1_{0,j}$ and $\psi^3_{l,m}$, 
 the first-order perturbation energy from $V_1+V^\dagger_1+V_2$, $E_{A,0}^{(1)}$ can be easily obtained
	\begin{align}
%	E_{A,0}^{(1)}=
%	\begin{cases}
     E^{(1)}_{A,n_1=0} = g^2v^2/2%~~\text{for}~~\psi^1_{0,j}
	, ~~~ %\\[2mm]
     E^{(1)}_{A,3} = g^2v^2.%~~\text{for}~~\psi^3_{l,m}. 
%	\end{cases}
	\end{align}
%and the degeneracy of $\psi^1_{0,j}$ and $\psi^3_{l,m}$ has been resolved.
Note that we have to solve secular equation for $\psi^1_{0,j}$ and $\psi^3_{l=0,m=0}$ 
 because there exist the degeneracy and the perturbation of $\psi^3_{l\ne0, m\ne0}$ can be obtained by $V_2$.
The second-order perturbation energy $E_{A,0}^{(2)}$ for $\psi^1_{0,j}$ is shown in appendix \ref{2ndpertur}.

For $\psi^1_{n+1,j},\psi^2_{n,j}~(n\geq 0)$ and $\psi^3_{l,m}$, the first-order perturbation energy 
 from $V_1+V^\dagger_1+V_2$, $E_{A}^{(1)}$ can be easily obtained
	\begin{align}
	E_{A,1^\prime}^{(1)}=0,~~~
	E_{A,2^\prime}^{(1)}=g^2v^2,%~~~~~\text{for}~~\psi^{2^\prime}_{n,j}
	\end{align}
where the mode functions in new direction $1^\prime$ and $2^\prime$ are defined as
%$\psi^{1^\prime}_{n+1,j}$ and $\psi^{2^\prime}_{n+1,j}$ are defined as
	\begin{align}
	\psi^{1^\prime}_{n+1,j}\equiv\left(i\psi^1_{n+1,j}+\psi^2_{n,j}\right)/\sqrt{2},~~~
	\psi^{2^\prime}_{n+1,j}\equiv\left(\psi^1_{n+1,j}+i\psi^2_{n,j}\right)/\sqrt{2}.
	\end{align}
The second-order perturbation energy $E_{A,1^\prime}^{(2)}$, $E_{A,2^\prime}$ and $E_{A,3}^{(2)}$ are shown in appendix \ref{2ndpertur}.

Thus, we summarize the mass of the gauge fields as
	\begin{align}
	\begin{cases}
	~m^2_{A,n_1=0}=\displaystyle\frac{g^2v^2}{2}+E_{A,0}^{(2)}, \\[6mm]
	~m^2_{A,1^\prime}\hspace{4.7mm}=\displaystyle \alpha_2(n+1)
		+E_{A,1^\prime}^{(2)}, \\[6mm]
	~m^2_{A,2^\prime}\hspace{4.7mm}=\displaystyle \alpha_2(n+1)
		+g^2v^2+E_{A,2^\prime}^{(2)}, \\[6mm]
	~m^2_{A,3}\hspace{5.65mm}=\displaystyle4\pi^2(l^2+m^2)+g^2v^2+E_{A,3}^{(2)},
	\end{cases}
	\label{gaugebosonmass}
	\end{align}
and we find that all of the gauge fields have nonzero mass if $v \ne0$. 
Therefore, we conclude that the SU(2) gauge symmetry is completely broken. 
The fact that non-zero VEV $v$ is realized will be seen in the potential analysis.

%%%%%%%%%%%%%%%%%%%%%%
\subsection{The mass of the scalar fields $\varphi^a$}
\label{SU(2)p}
%%%%%%%%%%%%%%%%%%%%%%
%The parts which are naively relevant to the scalar mass are
The terms relevant to the scalar mass are
 	\begin{align}
	\label{massp}
	\mathcal{L}&\supset-\frac{1}{4}\left\{(\mathcal{D}\bar\varphi
	+\overline{\mathcal{D}}\varphi)^2-2\sqrt{2}g(\partial\braket{\bar{\varphi}^a}
	+\bar\partial\braket{\varphi^a})[\varphi,\bar\varphi]^a\right\}
	+\frac{1}{4}(\mathcal{D}\bar\varphi-\overline{\mathcal{D}}\varphi)^2\nonumber \\
%	&=-\frac{1}{2}\left[\mathcal{D}\bar\varphi\overline{\mathcal{D}}\varphi
%+\overline{\mathcal{D}}\varphi\mathcal{D}\bar\varphi-2gf[\varphi,\bar\varphi]^3\right]\nonumber \\
	&=-\overline{\widetilde{\varphi}}\left(H
		+gf\hspace{0.6mm}\text{diag}(1,-1,0)\right)\widetilde{\varphi},
	\end{align}
where 
	\begin{align}
	\overline{\varphi}^{a^\prime}U_3^{a^\prime a}
		\equiv\overline{\widetilde{\varphi}}^a
		\equiv\sum_{\{n_a\}}\overline{\widetilde{\varphi}}^a_{\{n_a\}}\psi^a_{\{n_a\}}, 
		\qquad
%	~~\text{and}~~
	\left(U^{-1}_3\right)^{aa^\prime}\varphi^{a^\prime}
		\equiv\widetilde{\varphi}^a
		\equiv\sum_{\{n_a\}}\widetilde{\varphi}^a_{\{n_a\}}\psi^a_{\{n_a\}}.
	\end{align}
Since the energy eigenvalues for $\psi^1_{n,j}$ and $\psi^2_{n,j}$ are degenerate, 
 we must solve the secular equation. 
We find the first-order perturbation energy from $V_1+V^\dag_1+V_2$ as
	\begin{align}
	E^{(1)}_{\varphi,1^{\prime\prime}}=0,~~~%\text{for}~~\psi^{1^{\prime\prime}}_{n,j} \\[2mm]
	E^{(1)}_{\varphi,2^{\prime\prime}}=g^2v^2,%~~~~~\text{for}~~\psi^{2^{\prime\prime}}_{n,j}
	\end{align}
where the mode functions in new direction $1^{\prime\prime}$ and $2^{\prime\prime}$ are defined as
%$\psi^{1^{\prime\prime}}_{n,j}$ and $\psi^{2^{\prime\prime}}_{n,j}$ are defined as
	\begin{align}
	\psi^{1^{\prime\prime}}_{n,j}\equiv\left(i\psi^1_{n,j}+\psi^2_{n,j}\right)/\sqrt{2},~~~
	\psi^{2^{\prime\prime}}_{n,j}\equiv\left(\psi^1_{n,j}+i\psi^2_{n,j}\right)/\sqrt{2}.
	\label{psiprime}
	\end{align}
The second-order perturbation energy $E_{\varphi,1^{\prime\prime}}^{(2)}$, 
 $E_{\varphi,2^{\prime\prime}}^{(2)}$ and $E_{\varphi,3}^{(2)}$ are shown in appendix \ref{2ndpertur}.

Thus, the mass of the scalar fields are obtained as
	\begin{align}
	\begin{cases}
	~\displaystyle m^2_{\varphi,1^{\prime\prime}}\hspace{1.1mm}
		= \alpha_2\left(n+\frac{1}{2}\right)+E_{\varphi,1^{\prime\prime}}^{(2)}, \\[4mm]
	~\displaystyle m^2_{\varphi,2^{\prime\prime}}\hspace{1.1mm}
		= \alpha_2\left(n+\frac{1}{2}\right)+
		g^2v^2+E_{\varphi,2^{\prime\prime}}^{(2)}, \\[4mm]
	~\displaystyle m^2_{\varphi,3}\hspace{2.8mm}=4\pi^2(l^2+m^2)+g^2v^2+E_{\varphi,3}^{(2)}.
	\end{cases}
	\end{align}

%%%%%%%%%%%%%%%%%%%%%%
\subsection{The mass of the ghost fields $c^a$}
\label{SU(2)c}
%%%%%%%%%%%%%%%%%%%%%%
The terms %part which is 
 relevant to the ghost mass is
	\begin{align}
	\mathcal{L}&\supset-\bar c^a(\mathcal{D}_m\mathcal{D}^{m})^{ab}
		c^b~~(m=5,6),
	\end{align}
where %The square of $\mathcal{D}$ is
	\begin{align}
	\label{D^2}
	&\mathcal{D}_m\mathcal{D}^{m}\nonumber \\
	&=-\left(-\mathcal{D}\overline{\mathcal{D}}\right)-\frac{1}{2}
		\left[\mathcal{D},\overline{\mathcal{D}}\right]\nonumber \\
	&=-U_3\left[H+\frac{1}{2}\left(\left[\mathcal{D}_{3,diag},
		\overline{\mathcal{D}}_{3,diag}\right]-V_1-V_1^\dagger
		+V_3+V_3^\dagger\right)\right]U_3^{-1}\nonumber \\
	&=-U_3\left[H_0+V_2+gf\text{diag}(1,-1,0)
		+V_4+V_4^\dagger\right]U_3^{-1}. 
	\end{align}
%where 
$V_3$ and $V_4$ are defined as 
	\begin{align}
	V_3\equiv U_3^{-1}VU_3\mathcal{D}_{3,diag}
	~~\text{and}~~
	V_4\equiv\frac{1}{2}(V_1+V_3).
	\end{align}
%The first-, the second- and the third terms of the equation \eqref{D^2} are the same as those of the scalar fields.
Note that the first three terms in the equation \eqref{D^2} are the same as those of the scalar fields.
%The sum of the terms which have only diagonal components is
%	\begin{align}
%	&H_0+\frac{1}{2}\cdot2gf\text{diag}(1,-1,0)\nonumber \\
%	&=\left[\begin{array}{ccc}
%	2gf(n_1+1/2) & 0 & 0 \\
%	0 & 2gf(n_2+1/2) & 0 \\
%	0 & 0 & 4\pi^2(l^2+m^2) \end{array}\right].
%	\end{align}
%This is equal to the equation \eqref{phidiag}, so we can use the discussion of the scalar fields.
As in previous section, we solve the secular equation and find the first-order perturbation energy from $V_2+V_4+V^\dag_4$ as
	\begin{align}
	E^{(1)}_{c,1^{\prime\prime}}=0,~~~%\text{for}~~\psi^{1^{\prime\prime}}_{n,j} \\[2mm]
	E^{(1)}_{c,2^{\prime\prime}}=g^2v^2.%~~~~~\text{for}~~\psi^{2^{\prime\prime}}_{n,j}
	\end{align}
The second-order perturbation energy $E_{c,1^{\prime\prime}}^{(2)}$, $E_{c,2^{\prime\prime}}^{(2)}$ 
 and $E_{c,3}^{(2)}$ are summarized in appendix \ref{2ndpertur}.

Thus, the mass of the ghost fields are obtained as
	\begin{align}
	\begin{cases}
	~\displaystyle m^2_{c,1^{\prime\prime}}\hspace{1.1mm}
		= \alpha_2\left(n+\frac{1}{2}\right)+E_{c,1^{\prime\prime}}^{(2)}, \\[4mm]
	~\displaystyle m^2_{c,2^{\prime\prime}}\hspace{1.1mm}
		= \alpha_2\left(n+\frac{1}{2}\right)+
		g^2v^2+E_{c,2^{\prime\prime}}^{(2)}, \\[4mm]
	~\displaystyle m^2_{c,3}\hspace{2.8mm}=4\pi^2(l^2+m^2)+g^2v^2+E_{c,3}^{(2)}.
	\end{cases}
	\end{align}
Note that the mass of ghost fields are the same as that of scalar mass at the first order in Feynman gauge $\xi=1$.
This fact greatly simplifies the potential analysis as will be discussed later.

%%%%%%%%%%%%%%%%%%%%%%
\subsection{The analysis of the effective potential}
%%%%%%%%%%%%%%%%%%%%%%
Since the potential of the constant WL scalar fields is not generated at tree level, 
 we have to calculate the one-loop effective potential by use of KK mass spectrum obtained in the previous subsections.  
%We have not calculated the integrals of the products of the mode functions yet, however, 
We will show that the constant WL scalar VEV can be nonzero and the gauge symmetry is broken.
%Since $V_1+V_1^\dagger$ is proportional to $gv$, the lower order of the unknown integrals are $g^2v^2$.
%Although it is possibility that the second-order perturbation energy are proportional 
%to $g^2v^2$ since $V_1+V_1^\dagger$ is proportional to $gv$,

In order to calculate the one-loop effective potential as general as possible, 
% the typical forms of the mass are thus summarized below.
 we parametrize the KK mass spectrum as follows.  
	\begin{align}
	m^2_0\equiv A g^2v^2,~~~%+\mathcal{O}(g^4v^4),
		%\nonumber \\[2mm]
	m^2_n\equiv \alpha(n+x)+B g^2v^2,~~~%\nonumber \\[2mm]
		%+\mathcal{O}(g^4v^4),\nonumber \\[2mm]
	m^2_{l,m}\equiv 4\pi^2(l^2+m^2)+C g^2v^2,
		%+\mathcal{O}(g^4v^4),
	\end{align}
where $x$ is 0, $1$ or $1/2$ depending on the fields under consideration. 
We suppose that $A,B$ and $C$ are positive constant.
$\alpha$ means $\alpha_2$ or $\alpha_3$, which will be defined in next section.
%For a while, we ignore the term $\mathcal{O}(g^4v^4)$.
%We will refer to higher order effect later.
Then, the typical forms of the one-loop effective potential can be written as
	\begin{align}
	\nu_0(A)&\equiv\int\frac{d^4p}{(2\pi)^4}
		\ln\left[p^2+A g^2v^2\right], \\[2mm]
	\nu_n(\alpha,B;x)&\equiv\sum_{n=0}^\infty\int\frac{d^4p}{(2\pi)^4}
		\ln\left[p^2+\alpha(n+x)
		+Bg^2v^2\right], \label{withfluxpt} \\
%	\end{align}
%and
%	\begin{align}
	\nu_{l,m}(C)&\equiv\sum_{l=-\infty}^\infty\sum_{m=-\infty}^\infty
		\int\frac{d^4p}{(2\pi)^4}
		\ln\left[p^2+4\pi^2(l^2+m^2)
		+C g^2v^2\right]. \label{wofluxpt}
	\end{align}
Note that the second order perturbations are neglected in these potentials. 
For an obvious reason, $\nu_0(A)$, $\nu_n(\alpha,B;x)$ and $\nu_{l,m}(C)$ will be referred 
 as one-loop effective potential of zero-mode type, with-flux type and without-flux type, respectively.

%%%%%%%%%%%%%%%%%%%%%%
\subsubsection{The zero-mode type $\nu_0$}
%%%%%%%%%%%%%%%%%%%%%%
First, we consider the effective potential of zero-mode type $\nu_0$.
	\begin{align*}
	\nu_0(A)&=-\int\frac{d^4p}{(2\pi)^4}\int_0^\infty\frac{dt}{t}
		e^{-\left(p^2+A g^2v^2\right)t}\nonumber \\
	&=-\frac{1}{16\pi^2}\int_0^\infty\frac{dt}{t^3}
		e^{-A g^2v^2t}, 
	\end{align*}
where Schwinger's proper time integral is introduced in the first line and
the momentum integral is performed in the second line. 
Obviously, this integral diverges at $t=0$, but we can extract a finite value from it.
We will propose the idea later and the regularized effective potential of zero-mode type $\nu_{reg,0}$ 
 is found as 
	\begin{align}
	\nu_{reg,0}(A)=
%	&-\frac{1}{16\pi^2}
%		\left[\hspace{0.4mm}\zeta^{(1,0)}(-2,\mathfrak{a}g^2v^2)
%		-\frac{\mathfrak{a}g^2v^2}{12}(2\ln\mathfrak{a}g^2v^2+1)\right.
%		\nonumber \\
%	&\hspace{1.8cm}\left.+\frac{(\mathfrak{a}g^2v^2)^2}{2}
%		\ln\mathfrak{a}g^2v^2
%		-\frac{(\mathfrak{a}g^2v^2)^3}{9}(3\ln\mathfrak{a}g^2v^2-1)\right]
%	-\frac{(\mathfrak{a}g^2v^2)^2}{16\pi^2}\left[\hspace{0.5mm}\zeta^\prime(-2)+\frac{1}{36}\right].
	\frac{(A g^2v^2)^2}{576\pi^2}
		\left(\frac{9\hspace{0.5mm}\zeta(3)}{\pi^2}-1\right),
		\label{zeroreg}
	\end{align}
where $\zeta(3)$ is the Ap\'ery's constant, $\zeta(3)=1.20205\cdots$ and $9\hspace{0.5mm}\zeta(3)/\pi^2-1=0.0961381\cdots$.

%%%%%%%%%%%%%%%%%%%%%%
\subsubsection{The with-flux type $\nu_n$}
\label{nun}
%%%%%%%%%%%%%%%%%%%%%%
%Second, 
Next, we consider the effective potential of with-flux type $\nu_n$.
	\begin{align}
	\label{withflux}
	\nu_n(\alpha,B;x)
	&=-\sum_{n}\int\frac{d^4p}{(2\pi)^4}\int_0^\infty\frac{dt}{t}
		e^{-\left(p^2+\alpha(n+x)
		+B g^2v^2\right)t}\nonumber \\
	&=-\frac{1}{16\pi^2}\int_0^\infty\frac{dt}{t^{3}}
		\frac{e^{-\alpha(x+B g^2v^2/\alpha-1)t}}
		{e^{\alpha t}-1}\nonumber \\
	&=-\frac{\alpha^2}{16\pi^2}\int_0^\infty\frac{dy}{y^{1-(-2)}}
		\frac{e^{-(x+B g^2v^2/\alpha-1)y}}{e^y-1}.
	\end{align}
To calculate this, we try to apply the integral representation of the Hurwitz $\zeta$ function \cite{zetafunction}
%この文献を引用するかは確認、議論すべきかも
	\begin{align}
	\label{Hurwitz}
	\zeta(s,a)=&\hspace{0.4mm}\frac{a^{-s}}{2}+\frac{a^{1-s}}{s-1}
		+\sum_{k=1}^n\frac{B_{2k}}{(2k)!}
		\hspace{0.2mm}a^{1-s-2k}(s)_{2k-1}\nonumber \\
	&+\frac{1}{\Gamma(s)}\int_0^\infty dy\hspace{0.2mm}
		\frac{e^{-ay}}{y^{1-s}}\left(\frac{1}{e^y-1}-\frac{1}{y}+\frac{1}{2}
		-\sum_{k=1}^n\frac{B_{2k}}{(2k)!}\hspace{0.2mm}y^{2k-1}\right),
	\end{align}
where $n$ is a non-negative integer, $B_n$ is the Bernoulli number and $(s)_n$ is the Pochhammer symbol.
The equation \eqref{Hurwitz} is satisfied with the conditions
	\begin{align}
	\label{condition}
	%\mathfrak{R}
	{\rm Re}~s>-(2n+1)~~(n\in\mathbb{Z},~n\geq0),~~s\neq1,~~
		%\mathfrak{R}
		{\rm Re}~a>0.
	\end{align}
Comparing the integral in \eqref{withflux} with the expression \eqref{Hurwitz}, 
we need to consider a case where $s=-2$.
Therefore, it is enough to take $n=1$ as follows:
	\begin{align}
	\label{Hurwitz1}
	\zeta(s,a)=&\hspace{0.4mm}\frac{a^{-s}}{2}+\frac{a^{1-s}}{s-1}+
		\frac{sa^{-s-1}}{12}\nonumber \\
	&+\frac{1}{\Gamma(s)}\int_0^\infty dy\hspace{1mm}\frac{e^{-ay}}{y^{1-s}}
		\left(\frac{1}{e^y-1}-\frac{1}{y}+\frac{1}{2}-\frac{y}{12}\right),
	\end{align}
where $(s)_1=s$ and $B_2=1/6$ are used. 
The integral which includes a term of $1/(e^y-1)$ corresponds to the with-flux type $\nu_n$.
On the other hand, we know the form of $\zeta(-2,a)$ with elementary functions:
	\begin{align}
	\label{zeta-2}
	\zeta(-2,a)
	=-\frac{1}{3}\sum_{k=0}^3{}_3\mathrm{C}_kB_{3-k}a^k
	=\frac{a^{-(-2)}}{2}+\frac{a^{1-(-2)}}{-2-1}+\frac{(-2)a^{-(-2)-1}}{12},
	\end{align}
and we find this to be equivalent to the first line of the equation \eqref{Hurwitz1}.

When we take $s=\epsilon-2~(\epsilon=(4-d)/2\ll1)$, 
 the equation \eqref{zeta-2} can be understood 
 as $\mathcal{O}(\epsilon^0)$ terms of the both sides of the equation \eqref{Hurwitz1}. 
Therefore, the integral in the second line of \eqref{Hurwitz1} is understood to be $\mathcal{O}(\epsilon)$.
Extracting the $\mathcal{O}(\epsilon)$ terms from the equation \eqref{Hurwitz1}, we obtain
	\begin{align}
	\label{int}
	&\int_0^\infty dy\hspace{1mm}\frac{e^{-ay}}{y^{3}}
		\left(\frac{1}{e^y-1}-\frac{1}{y}+\frac{1}{2}-\frac{y}{12}\right)
		\nonumber \\
	&=\frac{1}{2}\left[\zeta^{(1,0)}(-2,a)-\frac{a}{12}(2\ln a+1)
		+\frac{a^2}{2}\ln a-\frac{a^3}{9}(3\ln a-1)\right],
	\end{align}
where we used the expansion of $\zeta(\epsilon-2,a)$ in $\epsilon$:
	\begin{align}
	\zeta(\epsilon-2,a)=\zeta(-2,a)+\epsilon\zeta^{(1,0)}(-2,a)+\mathcal{O}(\epsilon^2).
	\end{align}
%The term $\zeta^{(1,0)}(-2,a)$ can be expanded around $a=0$ by the following series:
%	\begin{align}
%	\label{series}
%	\zeta^{(1,0)}(-2,a)=&-\frac{\zeta(3)}{4\pi^2}+a\left(\frac{1}{4}-2\ln G\right)
%		+\frac{a^2}{2!}\left(\frac{3}{2}-\ln 2\pi-2\ln a\right)\nonumber \\
%		&+\frac{a^3}{3!}(3-2\hspace{0.2mm}\gamma_E)
%		+2\sum_{m=1}^\infty\frac{(-a)^{m+3}\zeta(m+1)}
%		{(m+3)(m+2)(m+1)},
%	\end{align}
%where $G$ is the Glaisher constant. %この展開が必要か不明However 
Although each integral of the equation \eqref{int} diverges at $y=0$, the right hand side of \eqref{int} is finite.
In other words, we can interpret that the last three terms of the left hand side in \eqref{int} work 
 as the regulators which extract finite quantity from the divergent integral 
 and we are able to evaluate the potential of with-flux type $\nu_n$.
Then, we obtain the regularized one-loop effective potential of with-flux type $\nu_{reg,n}$
	\begin{align}
	\nu_{reg,n}(\alpha,B;x)&=-\frac{\alpha^2}{32\pi^2}\left[\hspace{0.5mm}\zeta^{(1,0)}
		\left(-2,x+\frac{Bg^2v^2}{\alpha}-1
		\right)\right.\nonumber \\
	&\hspace{2.1cm}-\frac{1}{12}\left(x+\frac{Bg^2v^2}{\alpha}-1\right)\left\{2\ln\left(x+\frac{Bg^2v^2}{\alpha}-1\right)+1\right\}\nonumber \\
	&\hspace{2.1cm}+\frac{1}{2}\left(x+\frac{Bg^2v^2}{\alpha}-1\right)^2\ln\left(x+\frac{Bg^2v^2}{\alpha}-1\right)\nonumber \\
	&\hspace{2.1cm}\left.-\frac{1}{9}\left(x+\frac{Bg^2v^2}{\alpha}-1\right)^3\left\{3\ln\left(x+\frac{Bg^2v^2}{\alpha}-1\right)-1\right\}\right].
	\label{fluxtype:x=1}
	\end{align}

Note that the quantity $a=x+B g^2v^2/\alpha-1$ of the equation \eqref{withflux} is not always positive: 
 if $a$ is zero or negative, it does not satisfy the condition \eqref{condition}.
This situation occur in the case $x=0$ or 1/2.
In that case, we separate the term of $n=0$:
	\begin{align}
	\nu_{reg,n}(\alpha,B;x<1)=&-\int\frac{d^4p}{(2\pi)^4}\int_0^\infty\frac{dt}{t}e^{-\left(p^2+\alpha x+B g^2v^2\right)t} 
	\nonumber \\
	&-\sum_{n=1}^\infty\int\frac{d^4p}{(2\pi)^4}\int_0^\infty\frac{dt}{t}
		e^{-\left(p^2+\alpha(n+x)+B g^2v^2\right)t} \nonumber \\
		=&-\frac{\alpha^2}{16\pi^2}\int_0^\infty\frac{dy}{y^3}e^{-(x+Bg^2v^2/\alpha)y}
		-\frac{\alpha^2}{16\pi^2}\int_0^\infty\frac{dy}{y^3}\frac{e^{-(x+B g^2v^2/\alpha)y}}{e^y-1}.
	\label{x<1regpt}
	\end{align}
%	\begin{align*}
%	-\frac{1}{16\pi^2}\int_0^\infty\frac{dt}{t^3}
%		e^{-(2gf\mathfrak{n}+\mathfrak{b}g^2v^2)t}.
%	\end{align*}
%This is the expansion of the zero-mode type $\nu_0$.
Although the first term in \eqref{x<1regpt} is similarly divergent at $y=0$, 
% the first term in \eqref{int} 
 it can be finite by using \eqref{int} as
%We use the previous discussion and obtain
	\begin{align*}
	-\frac{1}{16\pi^2}\int_0^\infty\frac{dt}{t^3}
		e^{-(\alpha x+Bg^2v^2)t}
%	=-\frac{(4\pi\mathfrak{n}N+\mathfrak{b}g^2v^2)^2}{16\pi^2}
%		\left[\hspace{0.5mm}\zeta^\prime(-2)+\frac{1}{36}\right].
	\rightarrow \frac{\alpha^2}{576\pi^2}\left(x+\frac{B g^2v^2}{\alpha}\right)^2
		\left(\frac{9\hspace{0.5mm}\zeta(3)}{\pi^2}-1\right).
	\end{align*}
%In the above operation, 
In this calculation, the terms except for the third term of the left hand side in \eqref{int} work as the regulators.
Thus, the regularized one-loop effective potential of with-flux type $\nu_{reg,n}(\alpha,B;x<1)$ can be obtained
	\begin{align}
	\nu_{reg,n}(\alpha,B;x<1)
	&=-\frac{\alpha^2}{32\pi^2}\left[\hspace{0.5mm}\zeta^{(1,0)}
		\left(-2,x+\frac{Bg^2v^2}{\alpha}
		\right)\right.\nonumber \\
	&\hspace{2.1cm}-\frac{1}{12}\left(x+\frac{Bg^2v^2}{\alpha}\right)\left\{2\ln\left(x+\frac{Bg^2v^2}{\alpha}\right)+1\right\}\nonumber \\
	&\hspace{2.1cm}+\frac{1}{18}\left(x+\frac{Bg^2v^2}{\alpha}\right)^2\left\{ 9\ln\left(x+\frac{Bg^2v^2}{\alpha}\right)-\frac{9\hspace{0.5mm}\zeta(3)}{\pi^2}+1\right\}\nonumber \\
	&\hspace{2.1cm}\left.-\frac{1}{9}\left(x+\frac{Bg^2v^2}{\alpha}\right)^3\left\{3\ln\left(x+\frac{Bg^2v^2}{\alpha}\right)-1\right\}\right].
	\label{fluxtype:x<1}
	\end{align}

%%%%%%%%%%%%%%%%%%%%%%
\subsubsection{The without-flux type $\nu_{l,m}$}
%%%%%%%%%%%%%%%%%%%%%%
Finally, we consider the one-loop effective potential of without-flux type $\nu_{l,m}$.
	\begin{align*}
	\nu_{l,m}(C)&=-\sum_{l=-\infty}^\infty\sum_{m=-\infty}^\infty\int_0^\infty \frac{dt}{t}\int\frac{d^4 p}{(2\pi)^4}e^{-\left[p^2+4\pi^2(l^2+m^2)+Cg^2v^2\right]t} \\
	&=-\frac{1}{16\pi^2}\sum_{l,m}\int_0^\infty \frac{dt}{t^3}e^{-\left[4\pi^2(l^2+m^2)+Cg^2v^2\right]t} \\
%	&=-\frac{1}{16\pi^2}\int_0^\infty \frac{dt}{t^3}e^{-Bv^2t}\sum_{l=-\infty}^\infty\sum_{m=-\infty}^\infty e^{-\beta(l^2+m^2)t} \\
	&=-\frac{1}{16\pi^2}\int_0^\infty du ue^{-Cg^2v^2/u}\sum_{l,m}e^{-4\pi^2(l^2+m^2)/u}.
	\end{align*}
Using the Poisson resummation formula,
	\begin{align}
	\sum_{l=-\infty}^\infty\exp\left[-\frac{4\pi^2(l+v)^2}{u}\right]=\sqrt{\frac{u}{4\pi}}
	\sum_{r=-\infty}^\infty e^{-ur^2/4\pm2\pi i rv} \label{Poissonresum}
	\end{align}
with $v=0$, we obtain
	\begin{align}
	\label{typelm}
	\nu_{l,m}(C)%&=-\frac{1}{16\pi^2}\int_0^\infty du ue^{-v^2/u}\frac{\pi u}{\beta}
	%\sum_{r=-\infty}^\infty\sum_{s=-\infty}^\infty e^{-\pi^2u(r^2+s^2)/\beta} \\
	=&-\frac{1}{64\pi^3}\int_0^\infty du u^2e^{-Cg^2v^2/u}\sum_{r=-\infty}^\infty\sum_{s=-\infty}^\infty e^{-u(r^2+s^2)/4}\nonumber \\
	=&-\frac{1}{64\pi^3}\sum_{r,s\neq0}\left(\frac{4}{r^2+s^2}\right)^3
	\int_0^\infty du u^2e^{-u}\exp\left[-\frac{Cg^2v^2(r^2+s^2)}{4u}\right]\nonumber \\[1mm]
	&+\frac{1}{64\pi^3}\int_0^\infty dt\hspace{0.5mm}\frac{e^{-Cg^2v^2t}}{t^3}\left(-\frac{1}{t}\right),
	\end{align}
where we note that a change of variable $t=(r^2+s^2)/u$ is performed in the second line of \eqref{typelm} except for $r=s=0$ mode.
For the first term of the equation \eqref{typelm}, we consider applying the modified Bessel function of second kind,
	\begin{align}
	K_\nu(z)&=\frac{1}{2}\left(\frac{z}{2}\right)^\nu\int_0^\infty dt t^{-\nu-1}\exp\left(-t-\frac{z^2}{4t}\right) \nonumber \\
	&=\frac{1}{2}\left(\frac{z}{2}\right)^{-\nu}\int_0^\infty du u^{\nu-1}\exp\left(-\frac{z^2}{4u}-u\right),
	\end{align}
where a change of variable $u=z^2/4t$ is performed in the second line. 
This is satisfied with the conditions %$\mathfrak{R}
${\rm Re}~\nu>-1/2,~|\arg z|<\pi/4$.
Thus, the first term of the equation \eqref{typelm} becomes
	\begin{align*}
	&-\frac{1}{64\pi^3}\sum_{r,s\neq0}\left(\frac{4}{r^2+s^2}\right)^3
	\int_0^\infty du u^2e^{-u}\exp\left[-\frac{Cg^2v^2(r^2+s^2)}{4u}\right] \\[1mm]
	&=-\frac{C^{3/2}g^3v^3}{4\pi^3}\sum_{r,s\neq0}\left(\frac{1}{r^2+s^2}\right)^{3/2} K_3\left(gv\sqrt{C(r^2+s^2)}\right).
	\end{align*}
The second term of the equation \eqref{typelm} is regularized by using the equation \eqref{int}:
	\begin{align*}
	&\frac{1}{64\pi^3}\int_0^\infty dt\hspace{0.5mm}\frac{e^{-Cg^2v^2t}}{t^3}\left(-\frac{1}{t}\right)
%	&=\frac{1}{64\pi^3}\cdot\frac{1}{2}\left[\hspace{0.5mm}\zeta^{(1,0)}(-2,Cg^2v^2)
%		-\frac{Cg^2v^2}{12}(2\lnCg^2v^2+1)\right.\\
%	&\hspace{2.4cm}\left.+\frac{(Cg^2v^2)^2}{2}\lnCg^2v^2
%		-\frac{(Cg^2v^2)^3}{9}(3\lnCg^2v^2-1)\right].
	\rightarrow\frac{(Cg^2v^2)^3}{4608\pi^3}\left(\frac{9\hspace{0.5mm}\zeta(3)}{\pi^2}-1\right).
	\end{align*}
%In the above operation, the first, the third and the fourth terms of the left hand side in \eqref{int} work as the regulators.
Therefore, the regularized one-loop effective potential of without-flux type is obtained 
	\begin{align}
	\nu_{reg,l,m}(C)=&-\frac{(Cg^2v^2)^{3/2}}{4\pi^3}\sum_{r,s\neq0}\left(\frac{1}{r^2+s^2}\right)^{3/2} 
	K_3\left(gv\sqrt{C(r^2+s^2)}\right)\nonumber \\
%	&+\frac{1}{64\pi^3}\cdot\frac{1}{2}\left[\hspace{0.5mm}\zeta^{(1,0)}(-2,Cg^2v^2)
%		-\frac{Cg^2v^2}{12}(2\lnCg^2v^2+1)\right.\nonumber \\
%	&\hspace{2.4cm}\left.+\frac{(Cg^2v^2)^2}{2}\lnCg^2v^2
%		-\frac{(Cg^2v^2)^3}{9}(3\lnCg^2v^2-1)\right].
	&+\frac{(Cg^2v^2)^3}{4608\pi^3}
		\left(\frac{9\hspace{0.5mm}\zeta(3)}{\pi^2}-1\right).
	\label{bulktype}
	\end{align}

%%%%%%%%%%%%%%%%%%%%%%
%\subsubsection{The pictures of the several types}
%%%%%%%%%%%%%%%%%%%%%%
	\begin{figure}[h]
	\begin{center}
	\includegraphics[scale=0.8]{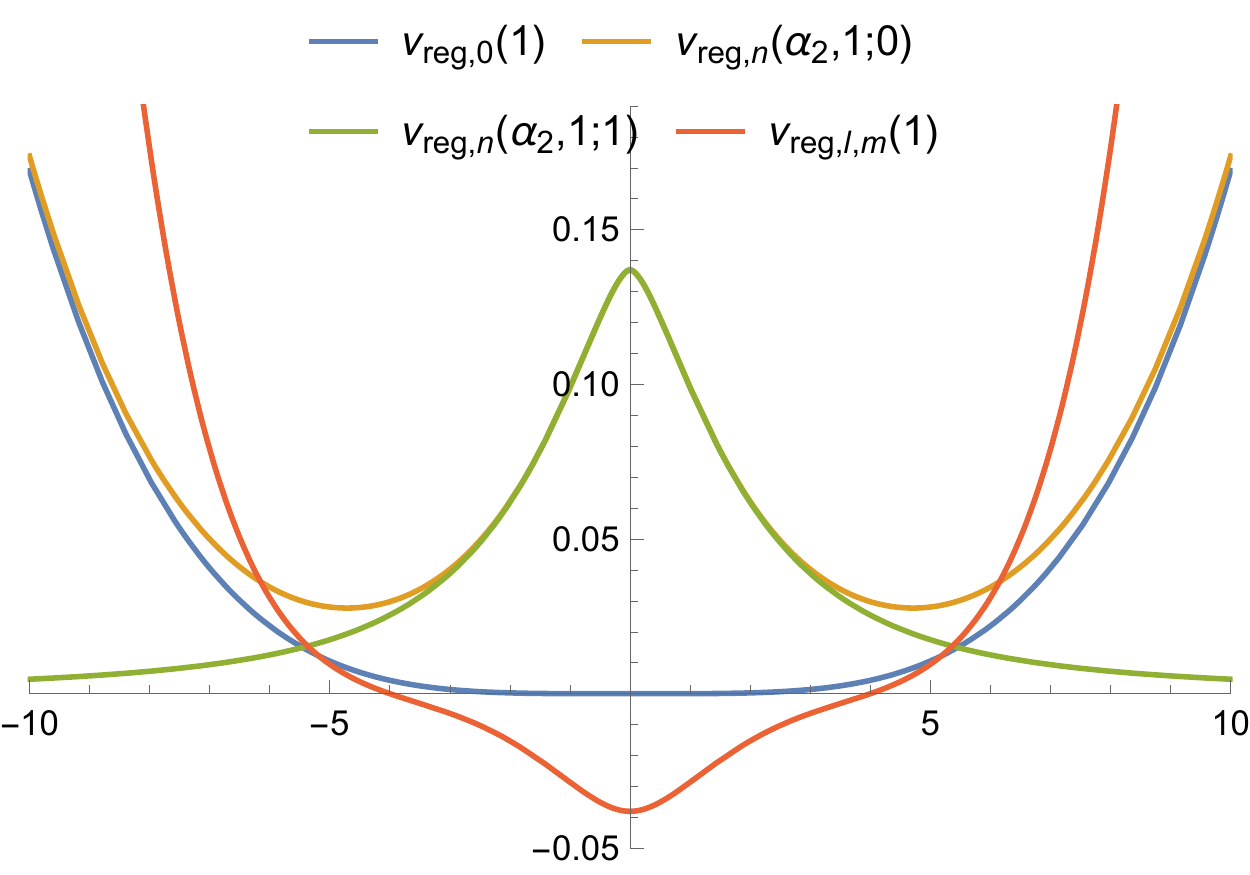}
	\caption{A picture of $\nu_{reg,0}(1)$, $\nu_{reg,n}(\alpha_2,1;0)$, $\nu_{reg,n}(\alpha_2,1;1)$ and $\nu_{reg,l,m}(1)$.}
	\label{several}
	\end{center}
	\end{figure}
%Finally, 
%we draw the pictures of t
The several types of the one-loop potentials $\nu_{reg,0}(1),\nu_{reg,n}(\alpha_2,1;0),\nu_{reg,n}(\alpha_2,1;1)$ 
 and $\nu_{reg,l,m}(1)$ are shown in Figure \ref{several}.

%%%%%%%%%%%%%%%%%%%%%%
\subsection{The one-loop effective potential of SU(2) Yang-Mills theory}
%%%%%%%%%%%%%%%%%%%%%%
We now apply the above results obtained in the previous subsections to SU(2) Yang-Mills theory.  
The effective potentials can be calculated by using the masses of the gauge fields $A^a_\mu$, 
 the scalar fields $\varphi^a$ and the ghost fields $c^a$,
	\begin{align}
	V_{A}&=\frac{3}{2}\frac{1}{(2\pi)^2}\Big(\nu_{0}(1/2)+\nu_{l,m}(1)+\nu_{n}(\alpha_2,1;1)\Big), \\
	V_{\varphi}&=\frac{1}{2}\frac{1}{(2\pi)^2}\Big(\nu_{l,m}(1)+\nu_{n}(\alpha_3,1;1/2)\Big), \\
	V_{c}&=-\frac{1}{2}\frac{1}{(2\pi)^2}\Big(\nu_{l,m}(1)+\nu_{n}(\alpha_3,1;1/2)\Big)=-V_{\varphi}.
	\end{align}
Since these effective potentials are divergent, we extract the finite value from them.
By using \eqref{zeroreg}, \eqref{fluxtype:x=1} and \eqref{bulktype}, the regularized effective potential is expressed as
	\begin{align}
	V_{reg,A}&=\frac{3}{2}\frac{1}{(2\pi)^2}\Big(\nu_{reg,0}(1/2)+\nu_{reg,l,m}(1)+\nu_{reg,n}(\alpha_2,1;1)\Big), \\
	V_{reg,\varphi}&=\frac{1}{2}\frac{1}{(2\pi)^2}\Big(\nu_{reg,l,m}(1)+\nu_{reg,n}(\alpha_3,1;1/2)\Big), \\
	V_{reg,c}&=-\frac{1}{2}\frac{1}{(2\pi)^2}\Big(\nu_{reg,l,m}(1)+\nu_{reg,n}(\alpha_3,1;1/2)\Big)=-V_{reg,\varphi}.
	\end{align}
Since $V_\varphi$ and $V_c$ cancel each other because of the same the KK mass spectrum in Feynman gauge,
 we have only to consider $V_{reg,A}$ as the total effective potential $V\equiv V_{reg,A}+V_{reg,\varphi}+V_{reg,c}$
\footnote{If you do not choose Feynman gauge $\xi=1$, the cancellation of $V_{reg,c}$ and $V_{reg,\varphi}$ does not occur.
%This implication has seen in \cite{HM}.
See \cite{HM} for this implications. 
}. %この注釈は一般的に成り立つわけではないことを一応伝えるために入れてる
In detail, the total effective potential can be expressed as
	\begin{align}
	V=&\hspace{1mm}V_{reg,A} \nonumber \\[1mm]
	=&\hspace{1mm}\frac{(g^2v^2)^2}{6144\pi^4}\left(\frac{9\zeta(3)}{\pi}-1\right) \nonumber \\
	&-\frac{3(g^2v^2)^{3/2}}{32\pi^5}\sum_{r,s\ne0}\left(\frac{1}{r^2+s^2}\right)^{3/2}K_3(gv\sqrt{r^2+s^2})
	+\frac{(g^2v^2)^3}{12288\pi^5}\left(\frac{9\zeta(3)}{\pi}-1\right) \nonumber \\
	&-\frac{3N^2}{16\pi^2}\left[\hspace{0.5mm}\zeta^{(1,0)}
	\left(-2,\frac{g^2v^2}{4\pi N}\right)\right.-\frac{1}{12}\left(\frac{g^2v^2}{4\pi N}\right)
	\left\{2\ln\left(\frac{g^2v^2}{4\pi N}\right)+1\right\}\nonumber \\
	&\hspace{1.7cm}+\frac{1}{2}\left(\frac{g^2v^2}{4\pi N}\right)^2\ln\left(\frac{g^2v^2}{4\pi N}\right)\left.
	-\frac{1}{9}\left(\frac{g^2v^2}{4\pi N}\right)^3\left\{3\ln\left(\frac{g^2v^2}{4\pi N}\right)-1\right\}\right].
	\end{align}

	\begin{figure}[h]
	\begin{center}
	\includegraphics[scale=0.8]{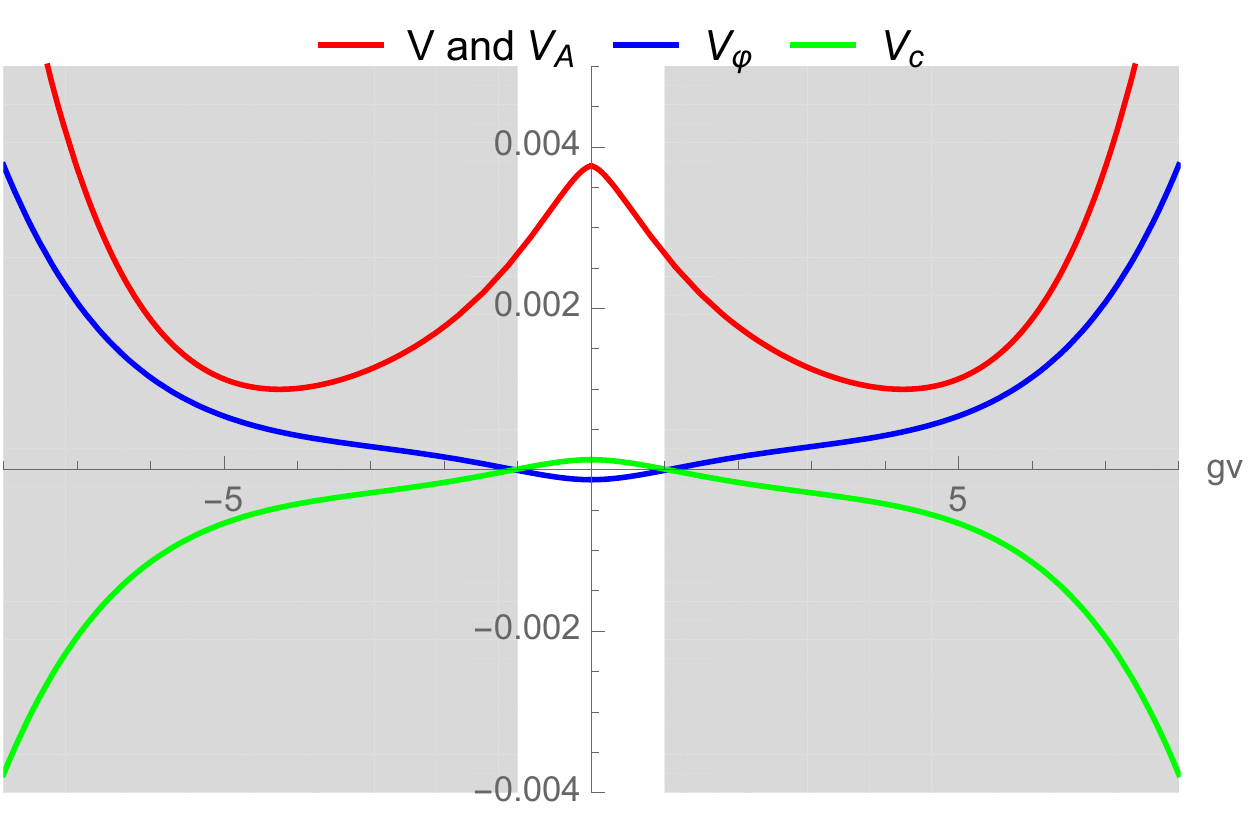}
	\caption{One-loop effective potentials of SU(2)Yang-Mills theory. 
	$N=3$ is taken for simplicity. 
	The shaded region represents a range where our perturbation analysis breaks down.}
	\label{effpt:sec2}
	\end{center}
	\end{figure}
The one-loop effective potentials with $N=3$ are shown in figure \ref{effpt:sec2}. 
The cancellation of the potentials between the scalar (blue line) and the ghost (green line) loop contributions can be explicitly verified.  
As can be seen from the total potential (red line) in figure \ref{effpt:sec2}, 
 we can conclude that the origin of the potential is at least not a minimum and 
 the minimum of the effective potential is expected to locate at a nonzero constant VEV $v\ne0$ 
 and the gauge symmetry SU(2) is completely broken.

Here we should note that we cannot determine the location of the minimum 
 since the location is expected to be beyond the range which our perturbation is valid as is shown in figure \ref{effpt:sec2}. 
Even if the value of the minimum is not determined, if we assume that the potential is bounded below, 
the pattern of gauge symmetry breaking can be confirmed from the gauge boson spectrum (\ref{gaugebosonmass}).
%One might think that our minimum is beyond the applicable range of perturbations $v \gg 1$, 
% but this is because the compactification radius is assumed to be $L=1$ for simplicity. 
%By adjusting $L$ as a free parameter, the minimum can be always in a range $v \ll 1$. 

%%%%%%%%%%%%%%%%%%%%%%
%\subsection{Higher order effect}
%%%%%%%%%%%%%%%%%%%%%%
%$\nu_{reg,0},\nu_{reg,n}$ and $\nu_{reg,l,m}$ are not likely to produce a local minimum at $gv\neq 0$, 
%but if we take higher order of $gv$, we might have a nonzero local minimum.
%To do that, we replace $g^2v^2$ by $g^2v^2-g^4v^4$, where the area $g^2v^2>1$ is forbidden.

%\newpage
%%%%%%%%%%%%%%%%%%%%%%
\section{SU(3) Yang-Mills theory %without perturbation
}%\label{SU(3)dir1,8}
%%%%%%%%%%%%%%%%%%%%%%
In a context of gauge-Higgs unification, 
 SU(3) gauge symmetry is minimal to realize the zero mode of WL scalar as an SU(2) Higgs doublet. 
In our case, if SU(3) is broken to SU(2) $\times$ U(1) by the VEV of the constant magnetic flux, 
 the SU(2) Higgs doublet would appear and the electroweak symmetry breaking can be discussed. 
As a first step toward the discussion, 
 we now consider a six-dimensional SU(3) Yang-Mills theory. %with two vacuum expectation values. 
%There are only two differences compared to the SU(2) theory: 
% ($\mathrm{i}$) gauge indices and ($\mathrm{ii}$) directions of VEVs.
%The gauge indices are $a,b,c=1,2,3,4,5,6,7,8$.
%Accordingly, the structure constants are changed to $f^{abc}$ of SU(3).
%We consider two types of directions where we introduce the VEVs:
We consider two cases of directions where we introduce the VEVs
	\begin{align}
	%\mathrm{ii}-\mathrm{i}.
	(1)~\braket{\phi^1}=\frac{v}{\sqrt{2}}~~\text{and}~~\braket{\phi^8}=\frac{f\bar z}{\sqrt{2}}, \\[4mm]
	%\mathrm{ii}-\mathrm{ii}.
	(2)~\braket{\phi^6}=\frac{v}{\sqrt{2}}~~\text{and}~~\braket{\phi^8}=\frac{f\bar z}{\sqrt{2}},
	\end{align}
where the superscripts of $\phi$ denote the gauge indices of SU(3) 
 and the structure constant is changed to that of SU(3) accordingly. 
%Other definitions which we discussed in the set-up section are all the same.
Note that the flux background breaks the SU(3) gauge symmetry, which is broken SU(2)$\times$ U(1).  %in SU(3) case.
The case (1) is not included in SU(2) theory, 
 because two kinds of VEV are both taken in the components of the unbroken symmetry. 
%Other definitions which we discussed in the set-up section are all the same.
Other cases of developing the VEVs are reduced to the above two cases by the gauge rotations. 
 
\subsection{Case (1)}
\label{SU(3)dir1,8}
In this subsection, let us consider the case (1) %$\mathrm{ii}-\mathrm{i}$ 
%which can be understood without perturbation theory.
%that will give us a useful suggestion.
%In this situation 
 where the background covariant derivatives $\mathcal D,\overline{\mathcal D}$ are expressed as
	\begin{align}
	\mathcal{D}^{ac}=\left[\begin{array}{cccccccc}
	\partial & 0 & 0 & 0 & 0 & 0 & 0 & 0 \\
	0 & \partial & igv & 0 & 0 & 0 & 0 & 0 \\
	0 & -igv & \partial & 0 & 0 & 0 &0 & 0 \\
	0 & 0 & 0 & \partial & \frac{\sqrt{3}}{2}igf\bar{z} & 0 & \frac{1}{2}igv & 0 \\
	0 & 0 & 0 & -\frac{\sqrt{3}}{2}igf\bar{z} & \partial & -\frac{1}{2}igv & 0 & 0 \\
	0 & 0 & 0 & 0 & \frac{1}{2}igv & \partial & \frac{\sqrt{3}}{2}igf\bar{z} & 0 \\
	0 & 0 & 0 & -\frac{1}{2}igv & 0 & -\frac{\sqrt{3}}{2}igf\bar{z} & \partial & 0 \\
	0 & 0 & 0 & 0 & 0 & 0 & 0 & \partial 
	\end{array}\right],
	\end{align}
	\begin{align}
	\overline{\mathcal{D}}^{ac}=\left[\begin{array}{cccccccc}
	\bar{\partial} & 0 & 0 & 0 & 0 & 0 & 0 & 0 \\
	0 & \bar{\partial} & -igv & 0 & 0 & 0 & 0 & 0 \\
	0 & igv & \bar{\partial} & 0 & 0 & 0 &0 & 0 \\
	0 & 0 & 0 & \bar{\partial} & -\frac{\sqrt{3}}{2}igfz & 0 & -\frac{1}{2}igv & 0 \\
	0 & 0 & 0 & \frac{\sqrt{3}}{2}igfz & \bar{\partial} & \frac{1}{2}igv & 0 & 0 \\
	0 & 0 & 0 & 0 & -\frac{1}{2}igv & \bar{\partial} & -\frac{\sqrt{3}}{2}igfz & 0 \\
	0 & 0 & 0 & \frac{1}{2}igv & 0 & \frac{\sqrt{3}}{2}igfz & \bar{\partial} & 0 \\
	0 & 0 & 0 & 0 & 0 & 0 & 0 & \bar{\partial} 
	\end{array}\right].
	\end{align}
Diagonalizing them, we obtain
	\begin{align}
	\begin{cases}
	\mathcal{D}_{diag}=\text{diag}\left(\partial,\partial-gv,\partial+gv,
	\partial-\frac{1}{2}gv-\frac{\sqrt{3}}{2}gf\bar z,\partial-\frac{1}{2}gv+\frac{\sqrt{3}}{2}gf\bar z,\right. \\[2mm]
     \hspace{2.5cm}\left.\partial+\frac{1}{2}gv-\frac{\sqrt{3}}{2}gf\bar z,\partial+\frac{1}{2}gv
     +\frac{\sqrt{3}}{2}gf\bar z,\partial\right), \\[5mm]
	\overline{\mathcal{D}}_{diag}
     =\text{diag}\left(\bar\partial,\bar\partial+gv,\bar\partial-gv,\bar\partial+\frac{1}{2}gv
     +\frac{\sqrt{3}}{2}gfz,\bar\partial+\frac{1}{2}gv-\frac{\sqrt{3}}{2}gfz,\right. \\[2mm]
     \hspace{2.5cm}\left.\bar\partial-\frac{1}{2}gv+\frac{\sqrt{3}}{2}gfz,\bar\partial-\frac{1}{2}gv-\frac{\sqrt{3}}{2}gfz,\bar\partial\right).
	\end{cases}
	\label{eigenvalue:sec3}
	\end{align}
Their diagonalizing unitary matrix $U$ is given by
	\begin{align}
	U=\frac{1}{2}\left[\begin{array}{cccccccc}
	2 & 0 & 0 & 0 & 0 & 0 & 0 & 0 \\
	0 & \sqrt{2} & \sqrt{2}i & 0 & 0 & 0 & 0 & 0 \\
	0 & \sqrt{2}i & \sqrt{2} & 0 & 0 & 0 &0 & 0 \\
	0 & 0 & 0 & 1 & i & -1 & i & 0 \\
	0 & 0 & 0 & i & 1 & -i & 1 & 0 \\
	0 & 0 & 0 & 1 & -i & 1 & i & 0 \\
	0 & 0 & 0 & i & -1 & i & 1 & 0 \\
	0 & 0 & 0 & 0 & 0 & 0 & 0 & 2
	\end{array}\right].
	\end{align}
The commutation relation of $\mathcal D$ and $\overline{\mathcal D}$ become just a constant matrix:
	\begin{align}
	\left[i\overline{\mathcal{D}}_{diag},i\mathcal{D}_{diag}\right]^{ac}
	=\sqrt{3}gf\text{diag}(0,0,0,1,-1,1,-1,0),
	%=\sqrt{3}gf\left[\begin{array}{cccccccc}
	%0 & 0 & 0 & 0 & 0 & 0 & 0 & 0 \\
	%0 & 0 & 0 & 0 & 0 & 0 & 0 & 0 \\
	%0 & 0 & 0 & 0 & 0 & 0 & 0 & 0 \\
	%0 & 0 & 0 & 1 & 0 & 0 & 0 & 0 \\
	%0 & 0 & 0 & 0 & -1 & 0 & 0 & 0 \\
	%0 & 0 & 0 & 0 & 0 & -1 & 0 & 0 \\
	%0 & 0 & 0 & 0 & 0 & 0 & 1 & 0 \\
	%0 & 0 & 0 & 0 & 0 & 0 & 0 & 0 \end{array}\right],
	%=-\sqrt{3}gf\left[a,a^\dagger\right]^{ac},
	\end{align}
where the creation and annihilation operators can be given as
	\begin{align}
	a=\frac{i}{\sqrt{\alpha_3}}\overline{\mathcal D}_{diag}, \qquad
	%~~\text{and}~~
	a^\dagger=\frac{i}{\sqrt{\alpha_3}}\mathcal D_{diag}~~(\alpha_3\equiv\sqrt{3}gf).
	\end{align}
Defining further 
	\begin{align}
	a_{4,6}&=\frac{i}{\sqrt{\alpha_3}}\left(\bar{\partial}+\frac{\sqrt{3}}{2}gfz \right),
	~~~a^\dag_{4,6}=\frac{i}{\sqrt{\alpha_3}}\left(\partial-\frac{\sqrt{3}}{2}gf \bar{z} \right), \\
	a_{5,7}&=\frac{i}{\sqrt{\alpha_3}}\left(\bar\partial-\frac{\sqrt{3}}{2}gfz\right),
	~~~a^\dag_{5,7}=\frac{i}{\sqrt{\alpha_3}}\left(\partial+\frac{\sqrt{3}}{2}gf\bar{z} \right)
	\end{align}
in the matrix form of the creation and annihilation operator,
the diagonalized part of $\mathcal{D}_{diag}$ and $\overline{\mathcal{D}}_{diag}$ can be expressed as
	\begin{align}
	\begin{cases}
	\displaystyle (i\overline{\mathcal{D}}_{diag})^{44}=\sqrt{\alpha_3}a_{4,6}+\frac{1}{2}igv, \\[4mm]
	\displaystyle (i\overline{\mathcal{D}}_{diag})^{55}=\sqrt{\alpha_3}a_{5,7}+\frac{1}{2}igv, \\[4mm]
	\displaystyle (i\overline{\mathcal{D}}_{diag})^{66}=\sqrt{\alpha_3}a_{4,6}-\frac{1}{2}igv, \\[4mm]
	\displaystyle (i\overline{\mathcal{D}}_{diag})^{77}=\sqrt{\alpha_3}a_{5,7}-\frac{1}{2}igv,
	\end{cases}
	~~\text{and}~~~
	\begin{cases}
	\displaystyle (i\mathcal{D}_{diag})^{44}=\sqrt{\alpha_3}a^\dag_{4,6}-\frac{1}{2}igv, \\[4mm]
	\displaystyle (i\mathcal{D}_{diag})^{55}=\sqrt{\alpha_3}a^\dag_{5,7}-\frac{1}{2}igv, \\[4mm]
	\displaystyle (i\mathcal{D}_{diag})^{66}=\sqrt{\alpha_3}a^\dag_{4,6}+\frac{1}{2}igv, \\[4mm]
	\displaystyle (i\mathcal{D}_{diag})^{77}=\sqrt{\alpha_3}a^\dag_{5,7}+\frac{1}{2}igv.
	\end{cases}
	\end{align}
Note that other components are just spatial derivatives and do not play any role of creation and annihilation operators.

%%%%%%%%%%%%%%%%%%%%%%
\subsubsection{The mass of the gauge fields $A^a_\mu$}
%%%%%%%%%%%%%%%%%%%%%%
In the same way as the calculations in subsection \ref{SU(2)A}, 
 we %consider $-\mathcal{D}_{diag}\overline{\mathcal D}_{diag}$ and 
 obtain the gauge mass matrix
	\begin{align}
     -\mathcal{D}_{diag}\overline{\mathcal D}_{diag}\equiv m^2_A,
	\end{align}
where the components are 
	\begin{align}
	\begin{cases}
	~(m^2_A)^{11}=4\pi^2(l_1^2+m_1^2), &
	\displaystyle~(m^2_A)^{22}=4\pi^2\left\{l_2^2+\left(m_2
		-\frac{gv}{2\pi}\right)^2\right\}, \\[4mm]
	\displaystyle~(m^2_A)^{33}=4\pi^2\left\{l_2^3+\left(m_3
		+\frac{gv}{2\pi}\right)^2\right\}, &
	\displaystyle~(m^2_A)^{44}=\alpha_3 n_4+\frac{1}{4}g^2v^2, \\[4mm]
	\displaystyle~(m^2_A)^{55}=\alpha_3 (n_5+1)+\frac{1}{4}g^2v^2, &
	\displaystyle~(m^2_A)^{66}=\alpha_3 n_6+\frac{1}{4}g^2v^2, \\[4mm]
	\displaystyle~(m^2_A)^{77}=\alpha_3 (n_7+1)+\frac{1}{4}g^2v^2, &
	~(m^2_A)^{88}=4\pi^2(l_8^2+m_8^2).
	\end{cases}
	\label{SU(3)gaugemass}
	\end{align}
From the massless modes in \eqref{SU(3)gaugemass}, 
 we find that the %SU(3)$\rightarrow$ 
 SU(2)$\times$ U(1) gauge symmetry is broken into U(1) $\times$ U(1) if $v\ne0$.

%%%%%%%%%%%%%%%%%%%%%%
\subsubsection{The mass of the scalar fields $\varphi^a$}
%%%%%%%%%%%%%%%%%%%%%%
%In the same way we calculated in subsection \ref{SU(2)p}, 
The scalar masses are calculated from the terms
% we consider  the sum of $2\left(-\mathcal{D}_{diag}\overline{\mathcal D}_{diag}\right)^{ac}$ 
% and $-4igf(U^{-1})^{a^\prime a}f^{8a^\prime c^\prime}U^{c^\prime c}$ and obtain the scalar mass matrix
	\begin{align}
	2\left(-\mathcal{D}_{diag}\overline{\mathcal D}_{diag}\right)^{ac}
	-4igf(U^{-1})^{a^\prime a}f^{8a^\prime c^\prime}U^{c^\prime c}
	\equiv m^2_\varphi,
	\end{align}
and the resulting KK mass spectrums are found as 
	\begin{align}
	\begin{cases}
	~(m^2_\varphi)^{11}=4\pi^2(l_1^2+m_1^2), &
	~(m^2_\varphi)^{22}=\displaystyle4\pi^2\left\{l_2^2+\left(m_2-\frac{gv}{2\pi}\right)^2\right\}, \\[4mm]
	~(m^2_\varphi)^{33}=\displaystyle4\pi^2\left\{l_2^3+\left(m_3+\frac{gv}{2\pi}\right)^2\right\}, &
	~(m^2_\varphi)^{44}\displaystyle=\alpha_3\left(n_4+\frac{1}{2}\right)+\frac{1}{4}g^2v^2, \\[4mm]
	~(m^2_\varphi)^{55}=\displaystyle\alpha_3\left(n_5+\frac{1}{2}\right)+\frac{1}{4}g^2v^2, &
	~(m^2_\varphi)^{66}=\displaystyle\alpha_3\left(n_6+\frac{1}{2}\right)+\frac{1}{4}g^2v^2, \\[4mm]
	~(m^2_\varphi)^{77}=\displaystyle\alpha_3\left(n_7+\frac{1}{2}\right)+\frac{1}{4}g^2v^2, &
	~(m^2_\varphi)^{88}=4\pi^2(l_8^2+m_8^2).
	\end{cases}.
	\end{align}

%%%%%%%%%%%%%%%%%%%%%%
\subsubsection{The mass of the ghost fields $c^a$}
%%%%%%%%%%%%%%%%%%%%%%
%In the same way we calculated in subsection \ref{SU(2)c}, 
% we consider the sum of $-\mathcal{D}_{diag}\overline{\mathcal D}_{diag}$ and $(\sqrt{3}gf/2)[a,a^\dagger]$ and obtain 
The ghost masses are calculated from the terms 
	\begin{align}
	-\mathcal{D}_{diag}\overline{\mathcal D}_{diag}
		+\frac{\alpha_3}{2}[a,a^\dagger]\equiv m^2_c,
	\end{align}
and the resulting KK spectrum are obtained 
	\begin{align}
	\begin{cases}
	\displaystyle~(m^2_c)^{11}=4\pi^2 (l^2_1+m^2_1), &
	\displaystyle~(m^2_c)^{22}=4\pi^2 \left\{l^2_2
		+\left(m_2-\frac{gv}{2\pi}\right)^2\right\}, \\[4mm]
	\displaystyle~(m^2_c)^{33}=4\pi^2 \left\{l^2_3
		+\left(m_3+\frac{gv}{2\pi}\right)^2\right\}, &
	\displaystyle~(m^2_c)^{44}=\alpha_3\left(n_4+\frac{1}{2}\right)
		+\frac{1}{4}g^2v^2, \\[4mm]
	\displaystyle~(m^2_c)^{55}=\alpha_3\left(n_5+\frac{1}{2}\right)
		+\frac{1}{4}g^2v^2, &
	\displaystyle~(m^2_c)^{66}=\alpha_3\left(n_6+\frac{1}{2}\right)
		+\frac{1}{4}g^2v^2, \\[4mm]
	\displaystyle~(m^2_c)^{77}=\alpha_3\left(n_7+\frac{1}{2}\right)
		+\frac{1}{4}g^2v^2, &
	\displaystyle~(m^2_c)^{88}=4\pi^2 (l^2_8+m^2_8). 
	\end{cases}
	\end{align}

%%%%%%%%%%%%%%%%%%%%%%
\subsubsection{The potential calculation}
%%%%%%%%%%%%%%%%%%%%%%
The effective potential can be calculated by using the masses of the gauge fields $A^a_\mu$, 
 the scalar fields $\varphi^a$, the ghost fields $c^a$ and \eqref{withfluxpt} and \eqref{wofluxpt}:
	\begin{align}
	V_A&=\frac{3}{2}\frac{1}{(2\pi)^2}\Big(2\nu_{l,m}(0)+\nu_{+}(gv/2\pi)
	+\nu_{-}(gv/2\pi)+2\nu_{n}(\alpha_3,1/4;0)+2\nu_{n}(\alpha_3,1/4;1)\Big), \\
	V_\varphi&=\frac{1}{2}\frac{1}{(2\pi)^2}\Big(2\nu_{l,m}(0)+\nu_{+}(gv/2\pi)+\nu_{-}(gv/2\pi)+4\nu_{n}(\alpha_3,1/4;1/2)\Big), \\
	V_c&=-\frac{1}{2}\frac{1}{(2\pi)^2}\Big(2\nu_{l,m}(0)+\nu_{+}(gv/2\pi)+\nu_{-}(gv/2\pi)+4\nu_{n}(\alpha_3,1/4;1/2)\Big)=-V_\varphi,
	\end{align}
where $\nu_\pm(V)$ is defined as
	\begin{align}
	\nu_\pm(V)&\equiv\sum_{l=-\infty}^\infty\sum_{m=-\infty}^\infty\int\frac{d^4p}{(2\pi)^4}\ln\left[p^2+4\pi^2\left\{l^2+\left(m\pm V\right)^2\right\}\right].
	\end{align}
The terms which do not contain the constant VEV $v$, $\nu_{l,m}(0)$, are irrelevant to determine the potential minimum.

%%%%%%%%%%%%%%%%%%%%%%
%\subsubsection{The another without-flux type $\nu_{\pm}$}
%%%%%%%%%%%%%%%%%%%%%%
We consider here new one-loop effective potentials of without-flux type $\nu_{\pm}$.
	\begin{align*}
	\nu_\pm(V)
	&=-\sum_{l,m}\int_0^\infty\frac{dt}{t}\int\frac{d^4p}{(2\pi)^4}
		\exp\left({-\left[p^2+4\pi^2\left\{l^2+\left(m\pm V\right)\right\}t\right]}\right)\nonumber \\
	&=-\frac{1}{16\pi^2}\sum_{l,m}\int_0^\infty\frac{dt}{t^3}
		\exp\left(-4\pi^2\left\{l^2+\left(m\pm V\right)\right\}t\right)\nonumber \\
	&=-\frac{1}{16\pi^2}\int_0^\infty duu\sum_{l,m}\exp\left(-4\pi^2
		\left\{l^2+\left(m\pm V\right)^2\right\}/u\right).
	\end{align*}
Using the Poisson resummation formula \eqref{Poissonresum},
%	\begin{align}
%	\sum_{m=-\infty}^\infty\exp\left(-4\pi^2\rho\left(m\pm\frac{gv}
%		{2\pi}\right)^2/u\right)
%	=\sum_{s=-\infty}^\infty\sqrt{\frac{u}{4\pi\rho}}\exp\left(
%		-\frac{us^2}{4\rho}\pm igvs\right),
%	\end{align}
$\nu_{\pm}$ become
	\begin{align*}
	\nu_\pm(V)
	=-\frac{1}{64\pi^3}\sum_{r=-\infty}^\infty\sum_{s=-\infty}^\infty
		e^{\pm 2\pi iVs}
		\int_0^\infty duu^2e^{-u(r^2+s^2)/4}.
	\end{align*}
Since the term with $s=0$ has no dependence on $v$, 
 it may be removed and we obtain the regularized one-loop effective potentials of without-flux type $\nu_{reg,\pm}$
	\begin{align}
	\nu_{reg,\pm}(V)
	&=-\frac{1}{64\pi^3}\sum_{r}\sum_{s\neq0}e^{\pm 2\pi iVs}
		\int_0^\infty duu^2e^{-u(r^2+s^2)/4}\nonumber \\
%	&=-\frac{1}{32\pi^3}\sum_{r}\sum_{s\neq0}e^{\pm igvs}
%		\left\{\frac{4}{r^2+s^2}\right\}^3\nonumber \\
	&=-\frac{4}{\pi^3}\sum_{r=-\infty}^\infty\sum_{s=1}^\infty
		\frac{1}{(r^2+s^2)^3}\cos\left(2\pi Vs\right).
	\label{GHUtype}
	\end{align}
Note that this periodic potential often appears in gauge-Higgs unification \cite{KLY, ABQ}.

By using \eqref{fluxtype:x=1}, \eqref{fluxtype:x<1}, \eqref{bulktype} and \eqref{GHUtype},
each regularized effective potential is expressed as
	\begin{align}
	V_{reg,A}&=\frac{3}{2}\frac{1}{(2\pi)^2}\Big(2\nu_{reg,l,m}(0)+2\nu_{reg,\pm}(gv/2\pi)%\nonumber \\
	%&\hspace{40mm}
	+2\nu_{reg,n}(\alpha_3,1/4;0)+2\nu_{reg,n}(\alpha_3,1/4;1)\Big), \\
	V_{reg,\varphi}&=\frac{1}{2}\frac{1}{(2\pi)^2}\Big(2\nu_{reg,l,m}(0)+2\nu_{reg,\pm}(gv/2\pi)+4\nu_{reg,n}(\alpha_3,1/4;1/2)\Big), \\
	V_{reg,c}&=-\frac{1}{2}\frac{1}{(2\pi)^2}\Big(2\nu_{reg,l,m}(0)+2\nu_{reg,\pm}(gv/2\pi)
	+4\nu_{reg,n}(\alpha_3,1/4;1/2)\Big)=-V_{reg,\varphi}.
	\end{align}
Because of $V_{reg,c}=-V_{reg,\varphi}$, 
 we consider $V_{reg,A}$ as the total effective potential $V\equiv V_{reg,A}+V_{reg,\varphi}+V_{reg,c}$.
%\footnote{If you do not choose Feynman gauge $\xi=1$, the cancellation of $V_{reg,c}$ and $=V_{reg,\varphi}$ does not occur.
%This implication has seen in \cite{HM}.
%}. %この注釈は一般的に成り立つわけではないことを一応伝えるために入れてる
In detail, the total one-loop effective potential can be expressed as
	\begin{align}
	V=V_{reg,A}&\supset
	-\frac{3}{\pi^5}\sum_{r=-\infty}^\infty\sum_{s=1}^\infty\frac{1}{(r^2+s^2)^3}\cos\left(gvs\right) \nonumber \\
	&\quad-\frac{9N^2}{16\pi^2}\left[\hspace{0.5mm}\zeta^{(1,0)}\left(-2,\frac{g^2v^2}{8\sqrt{3}\pi N}\right)\right.
		-\frac{1}{12}\left(\frac{g^2v^2}{8\sqrt{3}\pi N}\right)
		\left(2\ln\frac{g^2v^2}{8\sqrt{3}\pi N}+1\right)
		\nonumber \\
	&\hspace{2.3cm}+\frac{1}{36}\left(\frac{g^2v^2}{8\sqrt{3}\pi N}\right)^2
		\left\{18 \ln\frac{g^2v^2}{8\sqrt{3}\pi N}-\frac{9\zeta(3)}{2\pi^2}
		+1 \right\}\nonumber \\
	&\hspace{2.3cm}\left.-\frac{1}{9}\left(\frac{g^2v^2}{8\sqrt{3}\pi N}\right)^3
		\left(3\ln\frac{g^2v^2}{8\sqrt{3}\pi N}-1\right)\right].
	\label{totaleffpt:sec3}
	\end{align}
In \eqref{totaleffpt:sec3}, $2\nu_{reg,l,m}(0)$ term was dropped because it is irrelevant to finding the potential minimum.

	\begin{figure}[htbp]
	  \begin{minipage}[b]{0.5\linewidth}
	    \centering
	    \includegraphics[keepaspectratio, scale=0.6]{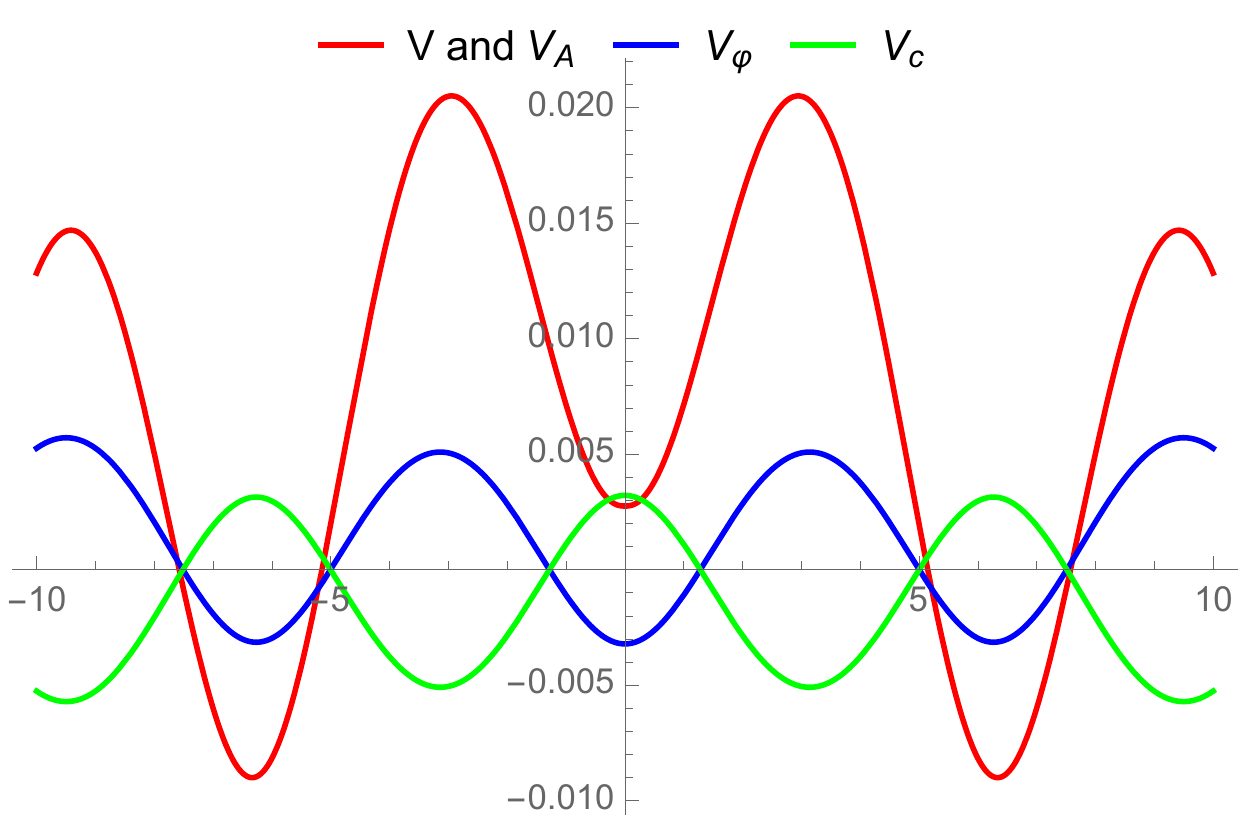}
	    \subcaption{The effective potential $V$.}
	  \end{minipage}
	  \begin{minipage}[b]{0.5\linewidth}
	    \centering
	    \includegraphics[keepaspectratio, scale=0.6]{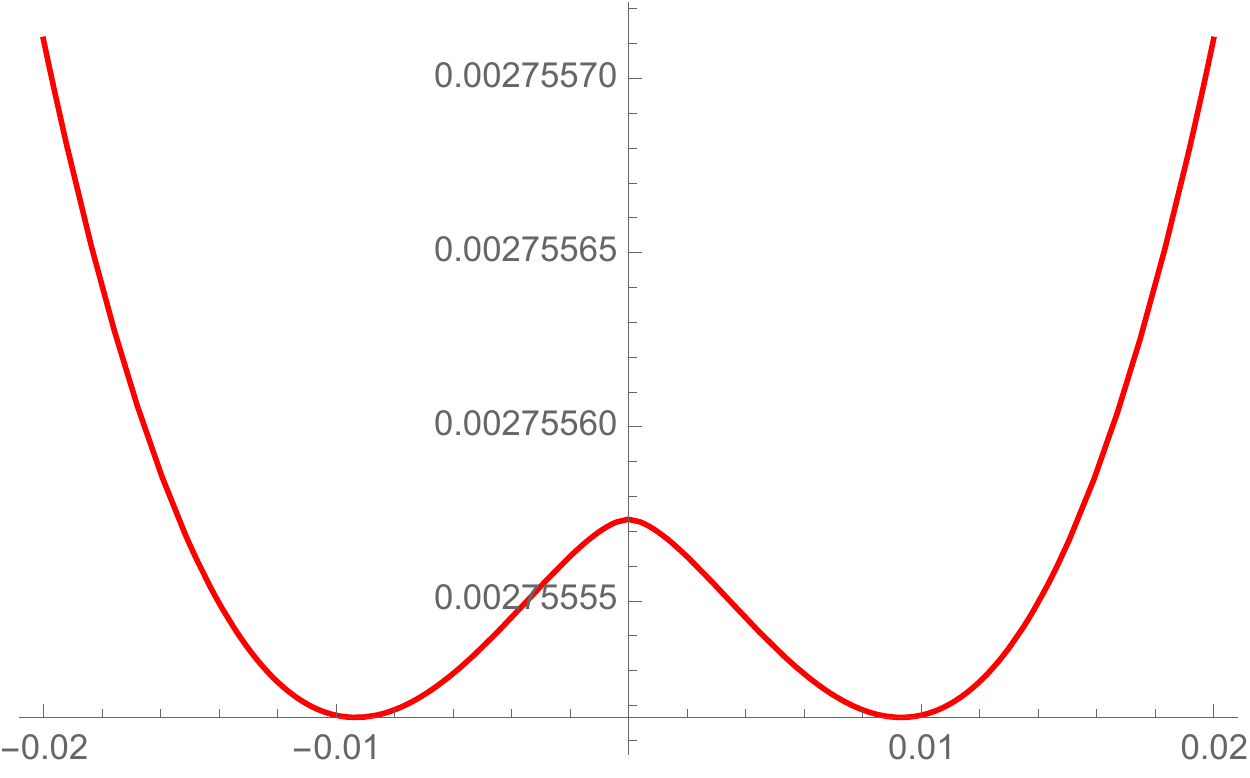}
	    \subcaption{A closeup of the effective potential.}
	  \end{minipage}
	  \caption{A picture of the effective potential $V$ with $N=3$ 
	  and its parts $V_A,V_\varphi~\text{and}~V_c$. (b) is a closeup of the effective potential (a).}
	  \label{effptfig:sec3}
	  \end{figure}

The effective potential \eqref{totaleffpt:sec3} of SU(3) Yang-Mills theory with %the Feynman gauge, $\xi=1$ and
$N=3$ is shown in Figure \ref{effptfig:sec3}.
%That is the particular case because $V_\varphi$ and $V_c$ cancel each other and $V_A$ itself becomes $V$.
Note that $V_\varphi$ and $V_c$ are cancelled each other and $V_A$ itself becomes the total potential $V$.
It seems that there is a local minimum at $gv=0$, but it is not correct.
The behavior of $V$ around $gv=0$ is shown in figure \ref{effptfig:sec3}(b) 
 and we find that there are two local minima at $gv\neq0$.
The reason why $V$ becomes convex upward at $gv=0$ is 
 due to the existence of the term $g^2v^2\ln g^2v^2$ in $V_A$: it becomes dominant as $gv$ gets close to zero. 
Furthermore, we emphasize that the logarithmic term is originated from the potential of with-flux type. 
If the magnetic flux is absent, we have no such contribution to the potential.  
Then, the potential is a periodic one as seen in the gauge-Higgs unification, 
 where the origin of the potential becomes convex downward, which implies the origin can be a local minimum. 
The effects from the potential of with-flux type are very crucial in our analysis of gauge symmetry breaking. 
Thus, we find that the minimum of the effective potential has a nonzero VEV $v\ne0$, %not $v=0$ 
and the gauge symmetry SU(2) $\times$ U(1) is broken into U(1) $\times$ U(1).

%\newpage
%%%%%%%%%%%%%%%%%%%%%%
%\section{SU(3) Yang-Mills theory with perturbation}
\subsection{Case (2)}\label{SU(3)dir6,8}
%%%%%%%%%%%%%%%%%%%%%%
In this section, we consider the case (2). %$\mathrm{ii}-\mathrm{ii}$,
	\begin{align}
	\braket{\phi^6}=\frac{v}{\sqrt{2}}, \qquad 
	  %~~\text{and}~~
	  \braket{\phi^8}=\frac{f\bar z}{\sqrt{2}}.
	\end{align}
%The direction of $\mathrm{ii}-\mathrm{ii}$ is assumed in gauge-Higgs unification.
In this case, the components of VEV where the constant WL scalar develops are 
 in the broken generators under SU(3) $\to$ SU(2) $\times$ U(1).
This case is similar to section \ref{SU(2)}.
In this situation, the background covariant derivatives $\mathcal{D}$, $\overline{\mathcal{D}}$ are expressed as
	\begin{align}
	\mathcal{D}^{ac}
	%&=\delta^{ac}\partial-\sqrt{2}igf^{abc}\braket{\phi^b}=\delta^{ac}\partial-\sqrt{2}igf^{a6c}\braket{\phi^6}-\sqrt{2}igf^{a8c}\braket{\phi^8} \\
	&=\left[\begin{array}{cccccccc}
	\partial & 0 & 0 & 0 & -\frac{1}{2}igv & 0 & 0 & 0 \\
	0 & \partial & 0 & \frac{1}{2}igv & 0 & 0 & 0 & 0 \\
	0 & 0 & \partial & 0 & 0 & 0 & \frac{1}{2}igv & 0 \\
	0 & -\frac{1}{2}igv & 0 & \partial & \frac{\sqrt{3}}{2}igf\bar{z} & 0 & 0 & 0 \\
	\frac{1}{2}igv & 0 & 0 & -\frac{\sqrt{3}}{2}igf\bar{z} & \partial & 0 & 0 & 0 \\
	0 & 0 & 0 & 0 & 0 & \partial & \frac{\sqrt{3}}{2}igf\bar{z} & 0 \\
	0 & 0 & -\frac{1}{2}igv & 0 & 0 & -\frac{\sqrt{3}}{2}igf\bar{z} & \partial & \frac{\sqrt{3}}{2}igv \\
	0 & 0 & 0 & 0 & 0 & 0 & -\frac{\sqrt{3}}{2}igv & \partial 
	\end{array}\right],
	\end{align}
	\begin{align}
	\overline{\mathcal{D}}^{ac}
	%&=\delta^{ac}\bar{\partial}+\sqrt{2}igf^{abc}\braket{\bar{\phi}^b}
	%=\delta^{ac}\bar{\partial}+\sqrt{2}igf^{a6c}\braket{\bar{\phi}^6}+\sqrt{2}igf^{a8c}\braket{\bar{\phi}^8} \\
	&=\left[\begin{array}{cccccccc}
	\bar{\partial} & 0 & 0 & 0 & \frac{1}{2}igv & 0 & 0 & 0 \\
	0 & \bar{\partial} & 0 & -\frac{1}{2}igv & 0 & 0 & 0 & 0 \\
	0 & 0 & \bar{\partial} & 0 & 0 & 0 & -\frac{1}{2}igv & 0 \\
	0 & \frac{1}{2}igv & 0 & \bar{\partial} & -\frac{\sqrt{3}}{2}igfz & 0 & 0 & 0 \\
	-\frac{1}{2}igv & 0 & 0 & \frac{\sqrt{3}}{2}igfz & \bar{\partial} & 0 & 0 & 0 \\
	0 & 0 & 0 & 0 & 0 & \bar{\partial} & -\frac{\sqrt{3}}{2}igfz & 0 \\
	0 & 0 & \frac{1}{2}igv & 0 & 0 & \frac{\sqrt{3}}{2}igfz & \bar{\partial} & -\frac{\sqrt{3}}{2}igv \\
	0 & 0 & 0 & 0 & 0 & 0 & \frac{\sqrt{3}}{2}igv & \bar{\partial} 
	\end{array}\right].
	\end{align}
If we diagonalize them, the eigenvalues are the same form as \eqref{diag:sec2}.
Therefore, we apply the perturbation theory as we have seen in section \ref{SU(2)A}.
Defining the unperturbed parts to be $\mathcal{D}_8$, $\overline{\mathcal{D}}_8$ 
 and the perturbation part $V$ such as \eqref{covSU(2)} and \eqref{perturSU(2)} respectively, 
 the covariant derivatives can be represented as  $\mathcal{D}=\mathcal{D}_8+V$ 
 and $\overline{\mathcal{D}}=\overline{\mathcal{D}}_8+\overline{V}$.
Diagonalizing $\mathcal{D}_8$, $\overline{\mathcal{D}}_8$, 
 we obtain the eigenvalues \eqref{eigenvalue:sec3} with $v=0$.
Their diagonalizing unitary matrix is given by
	\begin{align}
	U_8=\frac{1}{\sqrt{2}}\left[\begin{array}{cccccccc}
	\sqrt{2} & 0 & 0 & 0 & 0 & 0 & 0 & 0 \\
	0 & \sqrt{2} & 0 & 0 & 0 & 0 & 0 & 0 \\
	0 & 0 & \sqrt{2} & 0 & 0 & 0 &0 & 0 \\
	0 & 0 & 0 & 1 & i & 0 & 0 & 0 \\
	0 & 0 & 0 & i & 1 & 0 & 0 & 0 \\
	0 & 0 & 0 & 0 & 0 & 1 & i & 0 \\
	0 & 0 & 0 & 0 & 0 & i & 1 & 0 \\
	0 & 0 & 0 & 0 & 0 & 0 & 0 & \sqrt{2}
	\end{array}\right],
	\end{align}
which is different from $U$ in section \ref{SU(3)dir1,8}.
We can apply the discussion in section \ref{SU(2)A} by replacing $U_3$ in section \ref{SU(2)A} with $U_8$.
According to \eqref{DDbartransform}, we define $V_1\equiv \mathcal{D}_{8,diag} U^{-1}_8 V U_8$ and $V_2\equiv (U^{-1}_8 V U_8)^2$.
In particular, we focus on $V_2$:
	\begin{align}
	V_2=\frac{g^2v^2}{4}\left[\begin{array}{cccccccc}
	1 & 0 & 0 & 0 & 0 & 0 & 0 & 0 \\
	0 & 1 & 0 & 0 & 0 & 0 & 0 & 0 \\
	0 & 0 & 1 & 0 & 0 & 0 &0 & -\sqrt{3} \\
	0 & 0 & 0 & 1 & 0 & 0 & 0 & 0 \\
	0 & 0 & 0 & 0 & 1 & 0 & 0 & 0 \\
	0 & 0 & 0 & 0 & 0 & 2 & -2i & 0 \\
	0 & 0 & 0 & 0 & 0 & 2i & 2 & 0 \\
	0 & 0 & -\sqrt{3} & 0 & 0 & 0 & 0 & 3
	\end{array}\right].
	\end{align}
From \eqref{SU(3)gaugemass} with $v=0$ and $V_2$, 
 we find that the pair of $\psi^3_{l,m}$, $\psi^8_{l,m}$ and $\psi^6_{n+1,j}$, $\psi^7_{n,j}$ are degenerate,
and we must solve the secular equations.
As a result, the first-order perturbation energy $E^{(1)}$ has
	\begin{align}
	E^{(1)}_{3^\prime}=g^2v^2,~~~
	E^{(1)}_{8^\prime}=0,~~~
	E^{(1)}_{6^{\prime}}=0,~~~
	E^{(1)}_{7^{\prime}}=g^2v^2,
	\end{align}
where the mode functions in new direction $3^\prime$ and $8^\prime$ are defined as
	\begin{align}
	\psi^{3^\prime}_{l,m}&=(-\psi^3_{l,m}+\sqrt{3}\psi^{8}_{l,m})/2,~~~
	\psi^{8^\prime}_{l,m}=(\sqrt{3}\psi^3_{l,m}+\psi^{8}_{l,m})/2, \\
	\psi^{6^\prime}_{n,j}&=(i\psi^{6}_{n+1,j}+\psi^{7}_{n,j})/\sqrt{2},~~~
	\psi^{7^\prime}_{n,j}=(\psi^{6}_{n+1,j}+i\psi^{7}_{n,j})/\sqrt{2}.
	\end{align} 
Thus, the masses of gauge fields can be obtained as
	\begin{align}
	-\mathcal{D}_{diag}\overline{\mathcal D}_{diag}\equiv m^2_A,
	\end{align}
	\begin{align}
	\begin{cases}
	\displaystyle~(m^2_A)^{11}=4\pi^2(l_1^2+m_1^2)+\frac{1}{4}g^2v^2, &
	\displaystyle~(m^2_A)^{22}=4\pi^2(l_2^2+m_2^2)+\frac{1}{4}g^2v^2, \\[4mm]
	\displaystyle~(m^2_A)^{3^\prime 3^\prime}=4\pi^2(l_{3^\prime}^2+m^2_{3^\prime})+g^2v^2, &
	\displaystyle~(m^2_A)^{44}=\alpha_3 n_4+\frac{1}{4}g^2v^2, \\[4mm]
	\displaystyle~(m^2_A)^{55}=\alpha_3 (n_5+1)+\frac{1}{4}g^2v^2, &
	\displaystyle~(m^2_A)^{6^\prime 6^\prime}=\frac{1}{2}g^2v^2~(n_6=0),~\alpha_3(n_{6^\prime}+1)~(n_{6^\prime} \ge0),  \\[4mm]
	\displaystyle~(m^2_A)^{7^\prime 7^\prime}=\alpha_3 (n_{7^\prime}+1)+g^2v^2, &
	~(m^2_A)^{8^\prime 8^\prime}=4\pi^2(l_{8^\prime}^2+m_{8^\prime}^2),
	\end{cases}
	\label{SU(3)gaugemass:sec4}
	\end{align}
where we ignore the second order perturbation energy.

%\subsection{The effective potential}
	\begin{figure}[h]
	\begin{center}
	\includegraphics[scale=0.8]{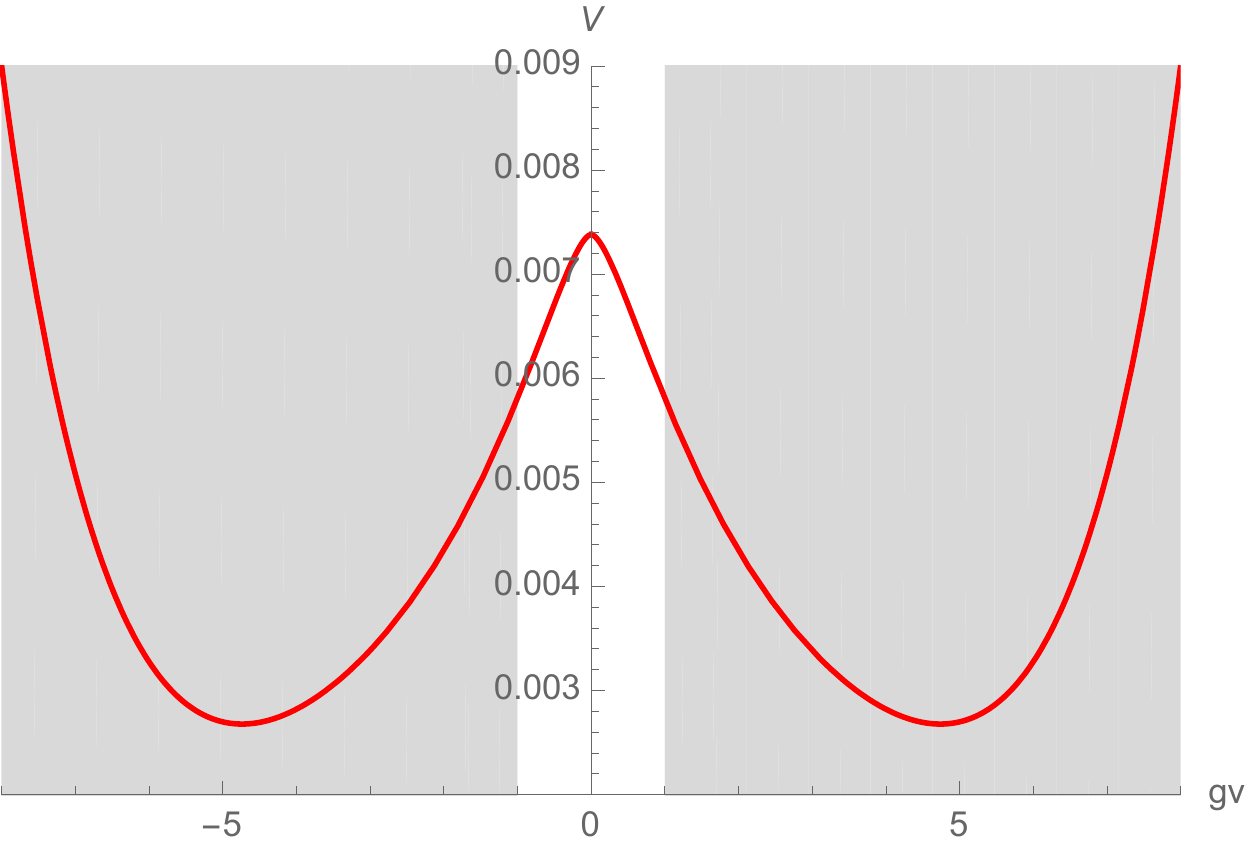}
	\caption{A picture of the effective potential with $N=3$.
	The shaded region represents a range where our perturbation analysis breaks down.}
	\label{effpt:sec4}
	\end{center}
	\end{figure}

In previous discussion, the total one-loop effective potential has been expressed by $V=V_{reg,A}$.
From \eqref{SU(3)gaugemass:sec4}, we have
	\begin{align}
	V&=\frac{3}{2}\frac{1}{(2\pi)^2}\Big(2\nu_{reg,l,m}(1/4)+\nu_{reg,l,m}(1)+\nu_{reg,n}(\alpha_3,1/4;0) \nonumber \\
	&\hspace{20mm}+\nu_{reg,n}(\alpha_3,1/4;1)+\nu_{reg,n}(\alpha_3,1;1)+\nu_{reg,0}(1/2)\Big).
	\end{align}
The effective potential of SU(3) Yang-Mills theory with $N=3$ is shown in Figure \ref{effpt:sec4}.
The shape of the potential in figure \ref{effpt:sec4} is similar to that (the red line) in Figure \ref{effpt:sec2}.
The minimum of the effective potential might have a nonzero constant VEV $v\ne0$ 
 and the gauge symmetry SU(2) $\times$ U(1) is broken to U(1).

We should comment here again as was done in SU(2) case 
 that we cannot determine the location of the minimum 
 since the location is expected to be beyond the range which our perturbation is valid 
 as is shown in figure \ref{effpt:sec4}. 
Even if the value of the minimum is not determined, if we assume that the potential is bounded below, 
the pattern of gauge symmetry breaking can be confirmed from the gauge boson spectrum (\ref{SU(3)gaugemass}).

%We comment that the minimum can be always in a range $v \ll 1$ by adjusting $L$ as in the SU(2) case. 

%It is instructive to compare our result with a higher dimensional theory with orbifold \cite{KLY}.
%余剰次元が違うもので引用しても大丈夫か？
%In SU(3) model in \cite{KLY}, SU(3) gauge symmetry is broken into SU(2)$\times$U(1) 
%because of orbifold, and a theory involves fermion fields.
%The effective potential with fermion fields can have a minimum at $v\ne0$,
%but the pattern of the gauge symmetry breaking is SU(2)$\times$U(1)$\rightarrow$U(1)$\times$U(1).
%On the other hand, although we do not introduce fermion field in our model, 
%the effective potential also has a minimum at $v\ne0$ and
%the pattern of the gauge symmetry breaking is SU(2)$\times$U(1)$\rightarrow$U(1) as we can see from \eqref{SU(3)gaugemass:sec4}.

%%%%%%%%%%%%%%%%%%%%%%
\section{Summary}
%%%%%%%%%%%%%%%%%%%%%%
We have studied six dimensional Yang-Mills theories compactified on a torus 
 with a magnetic flux and a constant VEV in this paper. 
Before constructing realistic models, 
 we have discussed simple models of SU(2) and SU(3) Yang-Mills theories 
 to understand basic structures of the gauge symmetry breaking. 

We first gave a setup of SU(2) model and derived the KK masses in terms of perturbation theory in quantum mechanics.
By using the KK masses, we calculated the one-loop effective potential.
In that computations, we focus on the integral representation of Hurwitz $\zeta$ function 
 and the regularized effective potential could be obtained.
From the obtained one-loop effective potential and the mass of the gauge fields, 
 we have seen that the SU(2) gauge symmetry is completely broken 
 because of the flux background and the constant VEV. 

Next, we considered an SU(3) model where 
%There are $N^2-1$ way of introduction of two VEVs in SU(N) gauge group,  and then We considered 
 two types of directions to introduce the flux background and the constant VEV. 
The extension to SU(3) is necessary for WL scalar fields to be an SU(2) doublet in the SM.  
In the case of section \ref{SU(3)dir1,8}, 
 the flux background and the constant VEV were introduced in the 8-th and the 1-st components of SU(3) gauge symmetry, respectively.  
% could compute the KK masses, 
The one-loop effective potential was calculated 
 and it was found that the potential has a nonzero VEV $v\ne0$, %, not $v=0$.In this case, 
 which implies that the SU(3) gauge symmetry is broken to U(1) $\times$ U(1) by the flux background and the constant VEV.
On the other hand, in the case of section \ref{SU(3)dir6,8}, 
 the flux background and the constant VEV were introduced in the 8-th and the 6-th components of SU(3) gauge symmetry, respectively.  
% which is assumed in gauge-Higgs unification, could not compute the KK masses, 
This case corresponds to gauge-Higgs unification in that the constant VEV is taken 
 in one of the component of the broken generators of the original symmetry, SU(3)/(SU(2)$\times$ U(1)) in our model. 
% and then we applied the KK masses in section \ref{SU(3)dir6,8} to the method of our SU(2) model.As in SU(2) model, 
The one-loop effective potential was found to expect a nonzero VEV $v\ne0$ 
 and the SU(3) gauge symmetry is broken to U(1) by the flux background and the constant VEV.
%Comparing our result with orbifold model \cite{KLY}, 
% the pattern of the gauge symmetry breaking SU(3)$\rightarrow$SU(2)$\times$U(1)$\rightarrow$U(1) is realized.

Although the results obtained in this paper are very interesting, it cannot be realistic as it stands. 
If we identify the WL scalar field with the SM SU(2) Higgs doublet in our SU(3) model, 
 the gauge symmetry breaking pattern SU(3) $\to$ SU(2) $\times$ U(1) $\to$ U(1)({\rm or}~U(1)$\times$ U(1)) 
 is not a correct pattern of the electroweak symmetry breaking SU(2) $\times$ U(1) $\to$ U(1). 
In order to realize such an electroweak symmetry breaking, 
 it would be interesting to take into account fermion field contributions to the one-loop effective potential.

In some models discussed in this paper, 
 the location of the potential minimum could not be determined in a parameter region where our perturbation is valid. 
It would be desirable to obtain the value of the potential minimum in a perturbative region 
 by extending our analysis to the models with fermions. 
These issues are left for our future work.

%%%%%%%%%%%%%%%
\section*{Acknowledgments}
%%%%%%%%%%%%%%%
This work is supported by JSPS KAKENHI Grant Number JP22J15562 (T.H.).
%This work is supported in part by JSPS KAKENHI Grant Number JP17K05420 (N.M.).

%%%%%%%%%%%%%%%%%%%%%%
\appendix
%%%%%%%%%%%%%%%%%%%%%
%%%%%%%%%%%%%%%%%%%%%
\section{The second-order perturbation energy}\label{2ndpertur}
%%%%%%%%%%%%%%%%%%%%%
In this appendix, we represent the second-order perturbation energy.
Since $V_2$ has an order of $\mathcal{O}(g^2v^2)$,
the second-order perturbation energy from $V_2$ has an order of $\mathcal{O}((g^2v^2)^2)$ and we neglect it.
\subsection{The mass of the gauge field}
%For $\psi^1_{0,j}$ and $\psi^3_{l,m}$, 
The second-order perturbation energy from $V_1+V^\dag_1$ for $\psi^1_{0,j}$ is
%$E_{A,0}^{(2)}$ for $\psi^1_{0,j}$ is
	\begin{align}
	E_{A,0}^{(2)}=&-\sum_{(l,m)\ne(0,0)}
		\frac{\left|\int_{T^2}d^2x\left(\psi^3_{l,m}\right)^\dagger
			(V_1+V^\dagger_1)\psi^1_{0,j}\right|^2}
		{4\pi^2(l^2+m^2)} \nonumber \\
%		-\sum_{n,j^\prime}\frac{\left|\int_{T^2}d^2x\left(\psi^2_{n,j^\prime}\right)
%		^\dagger V_2\psi^1_{0,j}\right|^2}{\alpha_2(n+1)}\nonumber \\
%	=&-\frac{g^2v^2}{2}\sum_{l,m}\frac{\left|\int_{T^2}d^2x
%		\bar\lambda_{l,m}(\partial-\bar\partial-gfz)\xi_{0,j}\right|^2}
%		{4\pi^2(l^2+m^2)+g^2v^2/2}\nonumber \\
	=&-\frac{g^2v^2}{2}\left[\hspace{0.5mm}\sum_{(l,m)\ne(0,0)}\frac{4\pi^2(l^2+m^2)}
		{4\pi^2(l^2+m^2)}\left|\int_{T^2}
		d^2x~\bar\lambda_{l,m}\xi_{0,j}\right|^2\right] \nonumber \\
	%&\hspace{1.7cm}\left.+\sum_{n,j^\prime}\frac{g^2v^2/2}{\alpha_2(n+1)}
	%	\left|\int_{T^2}d^2x~\xi_{n,j^\prime}\xi_{0,j}\right|^2\right]\nonumber \\
	\equiv&-\frac{g^2v^2}{2}\left[\hspace{0.5mm}\sum_{(l,m)\ne(0,0)}
		\left|C_{l,m,0,j}\right|^2\right]. 
	%+\sum_{n,j^\prime}\frac{(g^2v^2/2)\left|X_{n,0,j^\prime,j}\right|^2}
	%	{\alpha_2(n+1)}\right].
	\end{align}
We do not describe the integrals $C_{l,m,0,j}$ %and $X_{n,0,j^\prime,j}$ 
 in detail.

%For $\psi^1_{n+1,j},\psi^2_{n,j}~(n\geq 0)$ and $\psi^3_{l,m}$, 
%The first-order perturbation energy from $V^\prime_1+V_1^{\prime\hspace{0.4mm}\dagger}$, $E_1^{(1)}$ are zero and 
The second-order perturbation energy from $V_1+V^\dag_1$ for $\psi^{1^\prime}_{n+1,j},\psi^{2^\prime}_{n+1,j},\psi^3_{l=0,m=0}$ and $\psi^3_{l\neq0,m\neq0}$ are
%, $E_A^{(2)}$ for $\psi^{1^\prime}_{n+1,j},\psi^{2^\prime}_{n+1,j}$ and $\psi^3_{l,m}$ are
	\begin{align}
	E_{A,1^\prime}^{(2)}=
	-\sum_{l,m}\frac{\left|\int_{T^2}d^2x\left(\psi_{l,m}^3\right)^\dagger
		\left(V_1+V_1^{\dagger}\right)
		\psi_{n+1,j}^{1^\prime}\right|^2}
		{4\pi^2(l^2+m^2)- \alpha_2(n+1)},
	\end{align}
	\begin{align}
	E_{A,2^\prime}^{(2)}=
	-\sum_{l,m}\frac{\left|\int_{T^2}d^2x\left(\psi_{l,m}^3\right)^\dagger
		\left(V_1+V_1^{\dagger}\right)
		\psi_{n+1,j}^{2^\prime}\right|^2}
		{4\pi^2(l^2+m^2)- \alpha_2(n+1)},
	\end{align}
	\begin{align}
	E_{A,3,l=0,m=0}^{(2)}=\hspace{1mm}
	&\sum_{n,j}\frac{\left|\int_{T^2}d^2x\left(\psi_{l,m}^3\right)^\dagger
		\left(V_1+V_1^{\dagger}\right)
		\psi_{n+1,j}^{1^\prime}\right|^2}
		{4\pi^2(l^2+m^2)- \alpha_2(n+1)} \nonumber \\[2mm]
		&+\sum_{n,j}\frac{\left|\int_{T^2}d^2x\left(\psi_{l,m}^3\right)^\dagger
		\left(V_1+V_1^{\dagger}\right)
		\psi_{n+1,j}^{2^\prime}\right|^2}
		{4\pi^2(l^2+m^2)- \alpha_2(n+1)}
	\end{align}
and
	\begin{align}
	E_{A,3,l\neq0,m\neq0}^{(2)}=\hspace{1.4mm}
	&\frac{g^2v^2}{2}\sum_{j} \left|C_{l,m,0,j}\right|^2\nonumber %\\[2mm]
%	+\sum_{n,j^\prime}\frac{(g^2v^2/2)\left|X_{n,0,j^\prime,j}\right|^2}
%		{ \alpha_2(n+1)-g^2v^2/2}\right]&
+\sum_{n,j}\frac{\left|\int_{T^2}d^2x\left(\psi_{l,m}^3\right)^\dagger
		\left(V_1+V_1^{\dagger}\right)
		\psi_{n+1,j}^{1^\prime}\right|^2}
		{4\pi^2(l^2+m^2)- \alpha_2(n+1)} \nonumber \\[2mm]
		&+\sum_{n,j}\frac{\left|\int_{T^2}d^2x\left(\psi_{l,m}^3\right)^\dagger
		\left(V_1+V_1^{\dagger}\right)
		\psi_{n+1,j}^{2^\prime}\right|^2}
		{4\pi^2(l^2+m^2)- \alpha_2(n+1)}.
	\end{align}

%%%%%%%%%%%%%%%%%%%%%%%
\subsection{The mass of the scalar field}
%%%%%%%%%%%%%%%%%%%%%%%
The second-order perturbation energy from $V_1+V_1^{\dagger}$, $E_\varphi^{(2)}$ are
	\begin{align}
	E_{\varphi,1^{\prime\prime}}^{(2)}=-\sum_{l,m}\frac{\left|\int_{T^2}d^2x
		\left(\psi_{l,m}^3\right)^\dagger\left(V_1
		+V_1^{\dagger}\right)
		\psi_{n,j}^{1^{\prime\prime}}\right|^2}
		{4\pi^2(l^2+m^2)- \alpha_2(n+1/2)},
	\end{align}
	\begin{align}
	E_{\varphi,2^{\prime\prime}}^{(2)}=-\sum_{l,m}\frac{\left|\int_{T^2}d^2x
		\left(\psi_{l,m}^3\right)^\dagger\left(V_1
		+V_1^{\dagger}\right)
		\psi_{n,j}^{2^{\prime\prime}}\right|^2}
		{4\pi^2(l^2+m^2)- \alpha_2(n+1/2)}
	\end{align}
and
	\begin{align}
	E_{\varphi,3}^{(2)}=&~\sum_{n,j}\frac{\left|\int_{T^2}d^2x
		\left(\psi_{l,m}^3\right)^\dagger\left(V_1
		+V_1^{\dagger}\right)
		\psi_{n,j}^{1^{\prime\prime}}\right|^2}
		{4\pi^2(l^2+m^2)- \alpha_2(n+1/2)}
		\nonumber \\[2mm]
		&+\sum_{n,j}\frac{\left|\int_{T^2}d^2x
		\left(\psi_{l,m}^3\right)^\dagger\left(V_1
		+V_1^{\dagger}\right)
		\psi_{n,j}^{2^{\prime\prime}}\right|^2}
		{4\pi^2(l^2+m^2)- \alpha_2(n+1/2)}.
	\end{align}

\subsection{The mass of the ghost field}
The second-order perturbation energy by $V_4+V_4^{\dagger}$, $E_4^{(2)}$ are
	\begin{align}
	E_{c,1^{\prime\prime}}^{(2)}=-\sum_{l,m}\frac{\left|\int_{T^2}d^2x
		\left(\psi_{l,m}^3\right)^\dagger\left(V_4
		+V_4^{\dagger}\right)
		\psi_{n,j}^{1^{\prime\prime}}\right|^2}
		{4\pi^2(l^2+m^2)- \alpha_2(n+1/2)},
	\end{align}
	\begin{align}
	E_{c,2^{\prime\prime}}^{(2)}=-\sum_{l,m}\frac{\left|\int_{T^2}d^2x
		\left(\psi_{l,m}^3\right)^\dagger\left(V_4
		+V_4^{\dagger}\right)
		\psi_{n,j}^{2^{\prime\prime}}\right|^2}
		{4\pi^2(l^2+m^2)- \alpha_2(n+1/2)}
	\end{align}
and
	\begin{align}
	E_{c,3}^{(2)}=&~\sum_{n,j}\frac{\left|\int_{T^2}d^2x
		\left(\psi_{l,m}^3\right)^\dagger\left(V_4
		+V_4^{\dagger}\right)
		\psi_{n,j}^{1^{\prime\prime}}\right|^2}
		{4\pi^2(l^2+m^2)- \alpha_2(n+1/2)}
		\nonumber \\[2mm]
	&+\sum_{n,j}\frac{\left|\int_{T^2}d^2x
		\left(\psi_{l,m}^3\right)^\dagger\left(V_4
		+V_4^{\dagger}\right)
		\psi_{n,j}^{2^{\prime\prime}}\right|^2}
		{4\pi^2(l^2+m^2)- \alpha_2(n+1/2)}.
	\end{align}

 %%%%%%%%%%%


\begin{thebibliography}{100}
%%%%%%%%%%%%%%%

\bibitem{BKLS} 
  R.~Blumenhagen, B.~Kors, D.~Lust and S.~Stieberger,
  ``Four-dimensional String Compactifications with D-Branes, Orientifolds and Fluxes,''
  Phys.\ Rept.\  {\bf 445}, 1 (2007)
%  doi:10.1016/j.physrep.2007.04.003
  [hep-th/0610327].

\bibitem{IU}
 L. E. Ibanez and A. M. Uranga, 
 ``String theory and particle physics: An introduction to string theory," 
 Cambridge Univ. Press (2012).

\bibitem{Witten} 
E.~Witten,
``Some Properties of O(32) Superstrings,''
Phys.\ Lett.\  {\bf 149B}, 351 (1984).

\bibitem{ACKO}
H. Abe, K. S. Choi, T. Kobayashi and H. Ohki,
``Three generation magnetized orbifold models,"
Nucl. Phys. B 814, 265-292 (2009) [hep-th/0812.3534].

\bibitem{CIM}
D. Cremades, L. E. Ibanez and F. Marchesano, 
 ``Computing Yukawa couplings from magnetized extra dimensions," 
 JHEP \textbf{0405} 079 (2004) [hep-th/0404229]. 

  
\bibitem{0903}
  H. Abe, K. -S. Choi, T. Kobayashi and H. Ohki, 
 ``Higher Order Couplings in Magnetized Brane Models," 
 JHEP \textbf{0906} 080 (2009) [hep-th/0903.3800].

\bibitem{highermode}
Y. Hamada and T. Kobayashi,
“Massive Modes in Magnetized Brane Models,"
Prog. Theor. Phys. \textbf{128} (2012) 903 [hep-th/1207.6867].

\bibitem{MS}
  Y. Matsumoto and Y. Sakamura, 
 ``Yukawa couplings in 6D gauge-Higgs unification on $T^2/Z_N$ with magnetic fluxes,"
PTEP 2016 (2016) 5, 053B06 [hep-th/1602.01994].

\bibitem{B1}W. Buchmuller, M. Dierigl, E. Dudas and J. Schweizer, 
 ``Effective field theory for magnetic compactifications," 
  JHEP \textbf{1704} (2017) 052 [hep-th/1611.03798].

\bibitem{Lee}D. M. Ghilencea and H. M. Lee, 
 ``Wilson lines and UV sensitivity in magnetic compactifications," 
 JHEP \textbf{1706} 039 (2017) [hep-th/1703.10418].

\bibitem{B2}W. Buchmuller, M. Dierigl, E. Dudas, 
 ``Flux compactifications and naturalness," 
 JHEP \textbf{1808} 151 (2018) [hep-th/1804.07497].

\bibitem{HM}
T. Hirose and N. Maru,
“Cancellation of One-loop Corrections to Scalar Masses in Yang-Mills Theory with Flux Compactification,"
JHEP \textbf{1908} (2019) 054
[hep-th/1904.06028].

\bibitem{2-loop}
M. Honda and T. Shibasaki,
“Wilson-line Scalar as a Nambu-Goldstone Boson in Flux Compactifications and Higher-loop Corrections,”
JHEP \textbf{03} (2020) 031
[hep-th/1912.04581].

%下の自身の文献を引くか悩み中
%\bibitem{HDO}
%T. Hirose and N. Maru,
%“Cancellation of One-loop Corrections to Scalar Masses in Flux Compactification with Higher Dimensional Operators,"
%J. Phys. G 48 (2021) 5, 055005
%[hep-th/2012.03494].

\bibitem{finite}
T. Hirose and N. Maru,
“Nonvanishing Finite Scalar Mass in Flux Compactification,"
JHEP 06 (2021) 159
[hep-th/2104.01779].

\bibitem{Manton}
N. S. Manton,
``A New Six-Dimensional Approach to the Weinberg-Salam Model,"
Nucl. Phys. B 158 (1979) 141-153.

\bibitem{Hosotani1}
Y.Hosotani, 
``Dynamical Mass Generation by Compact Extra Dimension,"
Phys. Lett. \textbf{126B} (1983) 309.

\bibitem{Hosotani2}
Y.Hostonani,
``Dynamical of Non-integrable Phases and Gauge Symmetry Breaking,"
ANNALS \textbf{190}, (1989) 233-253

\bibitem{HIL}
H.~Hatanaka, T.~Inami and C.S.~Lim,
  ``The Gauge hierarchy problem and higher dimensional gauge theories,''
  Mod.\ Phys.\ Lett.\ A {\bf 13}, 2601 (1998);
%  doi:10.1142/S021773239800276X
  [hep-th/9805067].


\bibitem{LMH}
  C.S.~Lim, N.~Maru and K.~Hasegawa,
 ``Six dimensional gauge-Higgs unification with an extra space S**2 and the
  hierarchy problem,''
    J.\ Phys.\ Soc.\ Jap.\  {\bf 77}, 074101 (2008);
  arXiv:hep-th/0605180.

\bibitem{MaruYamashita}
N.~Maru and T.~Yamashita,
 ``Two-loop calculation of Higgs mass in gauge-Higgs unification:
  5D  massless QED compactified on S**1,''
  Nucl.\ Phys.\ B {\bf 754}, 127 (2006);
  [arXiv:hep-ph/0603237].

\bibitem{HMTY}
  Y.~Hosotani, N.~Maru, K.~Takenaga and T.~Yamashita,
 ``Two loop finiteness of Higgs mass and potential in the gauge-Higgs unification,''
  Prog.\ Theor.\ Phys.\  {\bf 118}, 1053 (2007);
  [arXiv:0709.2844 [hep-ph]].

\bibitem{KLY}
Masahiro Kubo, C.S.Lim and Hiroyuki Yamashita,
``The Hosotani mechanism in bulk gauge theories with an orbifold extra space $S^1/Z_2$,"
Mod. Phys. Lett. A17 (2002) 2249; 
[hep-ph/0111327]

\bibitem{ABQ}
I.~Antoniadis, K.~Benakli and M.~Quiros,
``Finite Higgs mass without supersymmetry,''
  New J.\ Phys.\  {\bf 3}, 20 (2001);
  [arXiv:hep-th/0108005].

\bibitem{zetafunction}
E. Elizalde,
“Ten physical applications of spectral zeta functions, "
second edition, Berlin, Springer, Germany (2012)

\end{thebibliography}
\end{document}